\documentclass{aa}  

\usepackage[switch]{lineno}
\usepackage{graphicx}
\usepackage{natbib}

\usepackage{supertabular}

\usepackage{txfonts}

\usepackage[]{hyperref}

\usepackage[usenames]{color}

\usepackage{amssymb}
\usepackage{mathrsfs}

\begin{document} 


   \title{Photometric detection of internal gravity waves in upper main-sequence stars}

   \subtitle{II. Combined TESS photometry and high-resolution spectroscopy}

   \titlerunning{TESS observations of IGWs in massive stars}
   
   \author{D. M. Bowman \inst{1} 
          \and
          S. Burssens \inst{1}
          \and
          S.~Sim{\' o}n-D{\' i}az \inst{2,3}
          \and
          P.~V.~F.~Edelmann \inst{4} 
          \and
          T.~M.~Rogers \inst{5,6}
       	  \and
          L.~Horst \inst{7}
          \and
          F.~K.~R{\"o}pke \inst{7,8}
          \and
          C.~Aerts \inst{1,9,10}
          }

    \institute{ Institute of Astronomy, KU Leuven, Celestijnenlaan 200D, B-3001 Leuven, Belgium \\
              \email{dominic.bowman@kuleuven.be} 
         \and
		Instituto de Astrof{\' i}sica de Canarias, E-38200 La Laguna, Tenerife, Spain
	\and	
		Departamento de Astrof{\' i}sica, Universidad de La Laguna, E-38205 La Laguna, Tenerife, Spain
	\and
		X~Computational Physics (XCP) Division and Center for Theoretical Astrophysics (CTA), Los Alamos National Laboratory, Los Alamos, NM 87545, USA
        \and
		School of Mathematics, Statistics and Physics, Newcastle University, Newcastle-upon-Tyne NE1 7RU, UK
	\and
		Planetary Science Institute, Tucson, AZ 85721, USA
	\and
		Heidelberger Institut f{\" u}r Theoretische Studien, Schloss-Wolfsbrunnenweg 35, 69118 Heidelberg, Germany
	\and
		Zentrum f{\" u}r Astronomie der Universit{\" a}t Heidelberg, Institut f{\" u}r theoretische Astrophysik, Philosophenweg 12, 69120 Heidelberg, Germany	
	\and
		Department of Astrophysics, IMAPP, Radboud University Nijmegen, NL-6500 GL Nijmegen, The Netherlands
	\and
		Max Planck Institute for Astronomy, Koenigstuhl 17, D-69117 Heidelberg, Germany	
            }

   \date{Received 21 April 2020 / accepted 1 June 2020}

  
 
  \abstract
   {Massive stars are predicted to excite internal gravity waves (IGWs) by turbulent core convection and from turbulent pressure fluctuations in their near-surface layers. These IGWs are extremely efficient at transporting angular momentum and chemical species within stellar interiors, but they remain largely unconstrained observationally.}
   {We aim to characterise the photometric detection of IGWs across a large number of O and early-B stars in the Hertzsprung--Russell diagram, and explain the ubiquitous detection of stochastic variability in the photospheres of massive stars.}
   {We combined high-precision time-series photometry from the NASA {\it Transiting Exoplanet Survey Satellite} with high-resolution ground-based spectroscopy of 70 stars with spectral types O and B to probe the relationship between the photometric signatures of IGWs and parameters such as spectroscopic mass, luminosity, and macroturbulence.}
  {A relationship is found between the location of a star in the spectroscopic Hertzsprung--Russell diagram and the amplitudes and frequencies of stochastic photometric variability in the light curves of massive stars. Furthermore, the properties of the stochastic variability are statistically correlated with macroturbulent velocity broadening in the spectral lines of massive stars.}
   {The common ensemble morphology for the stochastic low-frequency variability detected in space photometry and its relationship to macroturbulence is strong evidence for IGWs in massive stars, since these types of waves are unique in providing the dominant tangential velocity field required to explain the observed spectroscopy.}


   \maketitle


\section{Introduction}
\label{subsection: intro}

The advent of time-series photometry from space telescopes in the past decade has revealed a wealth of information for stars born with a convective core that was not previously attainable by ground-based telescopes. In particular, asteroseismology -- the study of stellar structure and evolution by means of forward modelling stellar pulsation frequencies -- has greatly benefitted from the long-term, high-precision, and continuous light curves assembled by space missions such as CoRoT \citep{Auvergne2009}, Kepler/K2 \citep{Borucki2010, Koch2010, Howell2014}, and more recently the {\it Transiting Exoplanet Survey Satellite} (TESS; \citealt{Ricker2015}) mission. This is because various phenomena exist in massive stars that produce variability with periods between minutes and decades, which in turn makes asteroseismology of pulsation mode frequencies in massive stars challenging when using ground-based telescopes \citep{ASTERO_BOOK}. However, space telescopes have enabled massive star asteroseismology by revealing a diverse range of different variability mechanisms across a wide range of masses, evolutionary stages, and metallicity environments \citep{Buysschaert2015, Rami2018b, Bowman2019b, Pedersen2019a, Handler2019a, Burssens2020a*}.

Despite the challenges in obtaining the necessary data suitable for asteroseismology, early studies of massive stars revealed that stellar structure and evolution theory is discrepant with observations in terms of angular momentum transport \citep{Aerts2019b} and chemical mixing \citep{Aerts2020a**}. In particular, the convective core masses of massive main-sequence stars inferred through asteroseismic modelling are larger than predicted by current theoretical models \citep{Handler2006a, Briquet2007e, Briquet2011, Daszy2013b, Aerts2019a}. Similarly, asteroseismology of main-sequence intermediate-mass stars also demonstrates the need for larger core masses than predicted by evolutionary models \citep{Moravveji2015b, Moravveji2016b, Schmid2016a, Buysschaert2018c, Szewczuk2018a, Mombarg2019a}. This core-mass discrepancy for massive stars is also evident in the detailed analysis of eclipsing double-lined spectroscopic binary (SB2) systems, in which a larger amount of extra mixing in the near-core region is needed to explain the location of binary systems in the Hertzsprung--Russell (HR) diagram \citep{Guinan2000, Claret2016b, Claret2019a, Johnston2019b, Tkachenko2020a}.

The evolution of a star born with a convective core involves the complex interaction of different physical processes, which are largely unconstrained in stellar models and currently controlled by free parameters, such as the amount and shape of near-core mixing. The large uncertainties from these unknowns are compounded by uncertainties associated with interacting multiple systems \citep{Sana2012b, Langer2012}, rotation \citep{Ekstrom2012a, Chieffi2013}, metallicity \citep{Georgy2013c, Groh2019a}, and magnetic fields \citep{Keszthelyi2019, Keszthelyi2020a}. Together these phenomena impact how a massive star is formed, how it evolves, and ultimately determine its fate beyond the main sequence. Yet the interior rotation, mixing, and angular momentum transport mechanism for main sequence massive stars remain largely unconstrained \citep{Aerts2019b}.

There are several non-mutually exclusive variability mechanisms operating in massive stars. For early-type stars, stochastic low-frequency variability at the surface is commonly observed in photometry (see e.g. \citealt{Balona1992c, Buysschaert2015, Bowman2019b}). On the other hand, spectroscopic variability in the form of spectral line profile variations and variable macroturbulence\footnote{Conversely, microturbulence has a length scale much shorter than the mean free path of a photon, and is not to be confused with macroturbulence. See \citet{GRAY_BOOK} for a detailed discussion.} are also typical for massive stars (see e.g. \citealt{Howarth1997, Simon-Diaz2014a}). In their detailed study of how macroturbulence is related to stellar parameters, \citet{Simon-Diaz2017a} make the following two important conclusions: (i) between early- and late-B main sequence stars, there is diverse behaviour in terms of line-broadening mechanisms, which is likely attributed to the diverse pulsational behaviour in this mass range; and (ii) main-sequence O stars and B supergiants have macroturbulence as their dominant broadening mechanism, with most stars sharing a common broadening profile. Previously, macroturbulence has been linked to non-radial pulsations (e.g. \citealt{Lucy1976e}), which are excited by either the opacity mechanism and/or turbulent pressure fluctuations in stellar envelopes \citep{Aerts2009b, Grassitelli2015a}.

The commonly known form of pulsations in massive stars are coherent pulsation modes (i.e. standing waves) triggered by an opacity mechanism operating in the Z-bump associated with iron-peak elements \citep{Dziembowski1993e, Dziembowski1993f, Miglio2007a, Szewczuk2017a, Godart2017}. Such a heat-engine is able to excite low-radial order pressure (p) modes in stars more massive than approximately 8~M$_{\rm \odot}$ and high-radial order gravity (g) modes in stars more massive than approximately 3~M$_{\rm \odot}$. See \citet{Szewczuk2017a} for calculations of instability regions of p- and g-mode pulsations in early-type stars including rotation. However, new observations reveal a significant fraction of pulsating massive stars outside of predicted instability regions \citep{Burssens2020a*}, hence the overall picture of variability is far from complete.

Additionally, quasi-periodic variability caused by surface and/or wind inhomogeneities combined with rotation \citep{Moffat2008c, David-Uraz2017b, Aerts2018a, Simon-Diaz2018a, Krticka2018e}, and small-scale variability triggered by thin subsurface convection zones associated with local opacity enhancements \citep{Cantiello2009a, Cantiello2011e, Cantiello2019, Lecoanet2013a} can play a role in some massive stars. Subsurface convection is predicted to be more efficient towards later evolutionary phases, hence microturbulence is predicted to increase with increasing luminosity and decreasing surface gravity, the latter of which having been confirmed by observations \citep{Cantiello2009a, Tkachenko2020a}. Despite the subsurface convection zone associated with the iron bump being absent in main-sequence B stars within the mass range $3 \leq M \leq 7$~M$_{\odot}$ \citep{Cantiello2009a, Cantiello2019}, stochastic variability has been detected in Slowly Pulsating B (SPB) stars observed by the CoRoT and Kepler space missions \citep{Bowman2019a, Pedersen_PhD}.

Moreover, subsurface convection zones and stellar winds do not directly provide the large-scale tangential velocity field needed to explain macroturbulent broadening in massive stars \citep{GRAY_BOOK, Simon-Diaz2010b, Simon-Diaz2014a, Simon-Diaz2017a}. A combined radial-tangential broadening profile is typically adopted to reproduce observed spectroscopic line profiles (see e.g. \citealt{Simon-Diaz2014a}). Such a profile from combining rotational and pulsational broadening components is motivated by the fact that non-radial gravity-mode pulsations produce predominantly horizontal velocities in the line-forming region. Rotational and microturbulent broadening alone cannot accurately reproduce (the variability of) spectral lines in hot stars \citep{GRAY_BOOK, Aerts2009b}.

Recent 2D and 3D hydrodynamical simulations demonstrate that internal gravity waves (IGWs) generated at the interface of the convective core and radiative envelope are also expected for massive stars \citep{Rogers2013b, Edelmann2019a, Horst2020a*}. These IGWs are efficient at transporting angular momentum and chemical species within stellar interiors \citep{Rogers2015, Rogers2017c, Edelmann2019a}. The collective power of an entire spectrum of IGWs excited by turbulent core convection is predicted to produce stochastic low-frequency variability and a large tangential velocity field near the stellar surface \citep{Rogers2013b, Edelmann2019a, Horst2020a*}. Hence an ensemble of IGWs provides the required velocity field to explain macroturbulent velocity broadening in hot stars \citep{Aerts2015c}. Furthermore, stochastic low-frequency variability was recently detected in hundreds of massive stars between 3 and 50~M$_{\odot}$ by \citet{Bowman2019b} and inferred to be caused by IGWs because of its similar morphology to that predicted by hydrodynamical simulations.

Here we combine high-precision TESS photometry and high-resolution ground-based spectroscopy to probe the relationship between a star's variability, location in the HR~diagram, and its measured macroturbulent broadening. In Section~\ref{section: method} we discuss our sample selection criteria and methodology. In Section~\ref{section: results}, we test how the morphology of stochastic low-frequency variability depends on the parameters of a star. Finally, we discuss our results in Section~\ref{section: discussion}, and conclude in Section~\ref{section: conclusions}.


\section{Method}
\label{section: method}

We extend the methodology developed by \citet{Bowman2019a} of analysing time-series photometry to a much larger sample of 70 massive stars. Our sample is comprised of early-type stars with spectral types O and B which have high-precision TESS photometry and fundamental parameters available from high-resolution spectroscopy \citep{Burssens2020a*}.

	\subsection{Sample selection criteria}
	\label{subsection: target selection}
		
	 As demonstrated by \citet{Blomme2011b} and \citet{Bowman2019a}, the variability in massive stars spans a broad range in frequency, and is significant above the instrumental white noise level at frequencies as high as 100~d$^{-1}$ for some stars. Therefore the TESS 2-min cadence is essential to avoid amplitude suppression of high-frequency variability\footnote{TESS full frame image (FFI) data have a cadence of 30~min which causes significant amplitude suppression near integer multiples of the FFI sampling frequency, hence prevents an accurate determination of the high-frequency component of IGWs above $\nu \gtrsim 10$~d$^{-1}$.} introduced by long-cadence time series photometry \citep{Murphy_PhD, Bowman_BOOK}. We exclude stars for which the 2-min TESS light curves exhibit strong signatures of contamination or instrumental systematics as identified by \citet{Burssens2020a*}. We also exclude eclipsing binary (EB) systems as these may contain a significant ($\gtrsim 1\%$) light contribution from a secondary component, causing the photometric variability detected in the primary to be modulated by the light ratio during the binary phase, and this limits scientific inference. 
	
	The spectroscopic parameters of our sample are provided by \citet{Burssens2020a*} and include the effective temperature, $T_{\rm eff}$, spectroscopic luminosity, $\log_{10}(\mathscr{L}/\mathscr{L_{\odot}})$, projected surface rotational velocity, $v\,\sin\,i$, and macroturbulent broadening, $v_{\rm macro}$. These were derived from high-resolution spectra assembled by the IACOB \citep{Simon-Diaz2011d, Simon-Diaz2015c} and OWN \citep{Barba2010, Barba2014, Barba2017} surveys. Within the IACOB spectroscopic database, spectra of northern OB stars have been obtained with the HERMES spectrograph ($R \simeq$~85\,000) on the 1.2-m Mercator telescope \citep{Raskin2011}, and the FIES spectrograph ($R \simeq$~46\,000) mounted on the 2.6-m NOT telescope \citep{Telting2014a}, on La Palma. Whereas spectra of southern O stars assembled as part of the OWN survey were obtained with the FEROS spectrograph ($R \simeq$~48\,000) mounted on the ESO/MPG 2.2-m telescope at La Silla \citep{Kaufer1997b, Kaufer1999b}. The extraction of spectroscopic parameters followed the methodologies outlined by \citet{Simon-Diaz2014a, Holgado2018a, Castro_N_2018b}, and we refer the reader to \citet{Burssens2020a*} for further details.
	
	Our final sample consists of 70 early-type stars in the southern ecliptic hemisphere (TESS sectors 1--13). The spectral types and fundamental parameters determined from spectroscopy obtained from \citet{Burssens2020a*} are provided in Table~\ref{table: stars}. Within our sample, typical uncertainties for $\log_{10}({\rm T}_{\rm eff})$ range between 0.03 and 0.05~dex, and for $\log_{10}(\mathscr{L}/\mathscr{L_{\odot}})$ range between 0.15 and 0.20~dex (see \citealt{Simon-Diaz2017a, Holgado2018a, Holgado2020a*}).


	\subsection{Iterative pre-whitening of coherent pulsation modes}
	\label{subsection: prewhitening}
	
	We obtain the 2-min TESS light curves from the Mikulski Archive for Space Telescopes (MAST\footnote{MAST website: \url{https://archive.stsci.edu/}}). We use the pre-search data conditioning simple aperture photometry (PDCSAP) time series, and refer the reader to \citet{Jenkins2016b} for further details of the TESS data pipeline. We perform checks to validate the chosen aperture mask and contamination, convert the light curves into units of stellar magnitudes and perform additional detrending in the form of a low-order polynomial for each sector. From high-precision 2-min TESS light curves, we calculate amplitude spectra by means of a discrete Fourier transform (DFT; \citealt{Deeming1975, Kurtz1985b}).
	
	In addition to requiring high-cadence time-series photometry, it is also necessary to remove high-amplitude pulsation modes to detect and characterise the underlying stochastic low-frequency variability in early-type stars \citep{Degroote2009a, Degroote2012b, Bowman2019a}. Importantly, we do not discriminate on the mechanism by which these coherent pulsation modes are excited: the opacity mechanism \citep{Szewczuk2017a} and/or by core convection \citep{Edelmann2019a, Horst2020a*}. It is known that the excitation of pulsation modes driven by the opacity mechanism is very sensitive to the rotation and metallicity of a star, and the opacity of stellar models \citep{Miglio2007a, Szewczuk2017a, Burssens2020a*}. On the other hand, core convection is predicted to excite a broad spectrum of IGWs, including resonant pressure and gravity eigenmodes \citep{Edelmann2019a, Lecoanet2019a, Horst2020a*}.
	
	We use the standard approach in asteroseismology of early-type stars with coherent pulsations and perform iterative pre-whitening to identify significant frequencies in the light curves and remove them to produce a residual amplitude spectrum (see e.g. \citealt{Degroote2009a, Papics2012a, VanReeth2015a, Bowman_BOOK}). In this iterative process, all high-amplitude coherent pulsation modes are subtracted from the observed light curve using the cosinusoidal model:
	
	\begin{equation}
	\Delta m = A \cos(2\pi\nu(t - t_0)+\phi) ~ ,
	\label{equation: sinusoid}
	\end{equation}
	
	\noindent where $A$ is the amplitude, $\nu$ is the frequency, $\phi$ is the phase, $t$ is the time with respect to a zero-point $t_0$. Following \citet{Bowman2019a, Bowman2019b}, we employ the standard amplitude significance criterion in iterative pre-whitening, which defines significant frequencies having an amplitude signal-to-noise ratio (S/N) larger than four \citep{Breger1993b}. We note that not all stars have significant frequencies following this definition (see \citealt{Blomme2011b, Bowman2019a}).


	\subsection{Amplitude spectrum fitting}
	\label{subsection: fitting}
	
	After iteratively prewhitening any significant pulsation mode frequencies and/or frequencies that may represent harmonics of the rotation frequency of each star, we use the method developed by \citet{Bowman2019a} and \citet{Bowman2019b} to characterise the stochastic low-frequency variability in its residual amplitude spectrum. We utilise a Bayesian Markov chain Monte Carlo (MCMC) framework with the \texttt{Python} code \texttt{emcee} \citep{Foreman-Mackey2013}{} to fit the residual amplitude spectrum of a star with the function:
	
	\begin{equation}
	\alpha \left( \nu \right) = \frac{ \alpha_{0} } { 1 + \left( \frac{\nu}{\nu_{\rm char}} \right)^{\gamma}} + C_{\rm w} ~ ,	
	\label{equation: red noise}
	\end{equation}
	
	\noindent where $\alpha_{0}$ represents the amplitude at a frequency of zero, $\gamma$ is the logarithmic amplitude gradient, $\nu_{\rm char}$ is the characteristic frequency (i.e. the inverse of the characteristic timescale, $\tau$, of stochastic variability present in the light curve such that $\nu_{\rm char} = (2\pi\tau)^{-1}$), and $C_{\rm w}$ is a frequency-independent (i.e. white) noise term \citep{Blomme2011b, Bowman2019b}.
	
	Similarly to \citet{Bowman2019b}, we use non-informative (flat) priors and 128 parameter chains, and we fit the (residual) amplitude spectrum up to the TESS Nyquist frequency (i.e. $0.1 \leq \nu \leq 360.0$~d$^{-1}$) using the model given in Eq.~(\ref{equation: red noise}) for each star in our sample. At each iteration, a parameter chain is used to construct a model that is subject to a log-likelihood evaluation:
	
	\begin{equation}
	\ln\mathcal{L} \propto - \frac{1}{2} \sum_i \left( \frac{y_i - M(\Theta_i)}{\sigma_i} \right)^{2} ~ ,
	\label{equation: likelihood}
	\end{equation}
	
	\noindent where $\ln\mathcal{L}$ is the log-likelihood, $y_i$ are the data, $\sigma_i$ are their uncertainties, and $M_{\Theta}$ is the model with parameters $\Theta$. After burning the first few hundred iterations, convergence is confirmed for the subsequent $\sim$1000 iterations using the parameter variance criterion from \citet{Gelman1992}.
	
	The residual amplitude spectrum fit using Eq.~(\ref{equation: red noise}) provides the amplitude and dominant timescale(s) of the stochastic variability in the light curve of each massive star. Furthermore, the steepness of the frequency spectrum, $\gamma$, and at what frequency the profile turns over, $\nu_{\rm char}$, are important parameters to distinguish the source of the stochastic variability \citep{Bowman2019a}. Hydrodynamical simulations predict that an ensemble of IGWs generated by core convection produces an amplitude spectrum with $0.8 \leq \gamma \leq 3$ and variability up to 100~d$^{-1}$. Furthermore, the exact values of $\gamma$ and $\nu_{\rm char}$ depend on the mass and radius of the host star \citep{Rogers2013b, Edelmann2019a, Horst2020a*}. On the other hand, other sources of low-frequency variability such as small-scale waves generated by subsurface convection produce a steeper frequency spectrum ($\gamma \geq 3.25$; see \citealt{Couston2018b}).
	
	At very low frequencies, such as below 0.1~d$^{-1}$, the probing power of time series photometry is limited by the length of observations. Furthermore, instrumental systematics present in the light curve may dominate in the amplitude spectrum below 0.1~d$^{-1}$ since they correspond to variability with periods of order the length of the time series. In the case of TESS data, this corresponds to approximately 12~d (i.e. half of a single sector), which is why \citet{Bowman2019b} only characterised variability above 0.1~d$^{-1}$. In all massive stars studied by \citet{Bowman2019b} and those in our current TESS sample, the measured $\nu_{\rm char}$ parameter is significant at frequencies higher than 0.2~d$^{-1}$.


\section{Results}
\label{section: results}

	The amplitude spectrum of each massive star in our sample was fit using Eq.~(\ref{equation: red noise}). For all 70 stars in our sample, we provide the resultant fit parameters in Table~\ref{table: params} and their $1\sigma$ statistical uncertainties as determined from the converged parameter chains from our MCMC framework \citep{Bowman2019a}. The fitted logarithmic (residual) amplitude spectra for all stars are provided as figures in Appendix~\ref{section: appendix: figures}, so as to demonstrate the broad range of frequencies and amplitudes within our sample of massive stars.
	
\begin{figure}
	\centering
	\includegraphics[width=0.99\columnwidth]{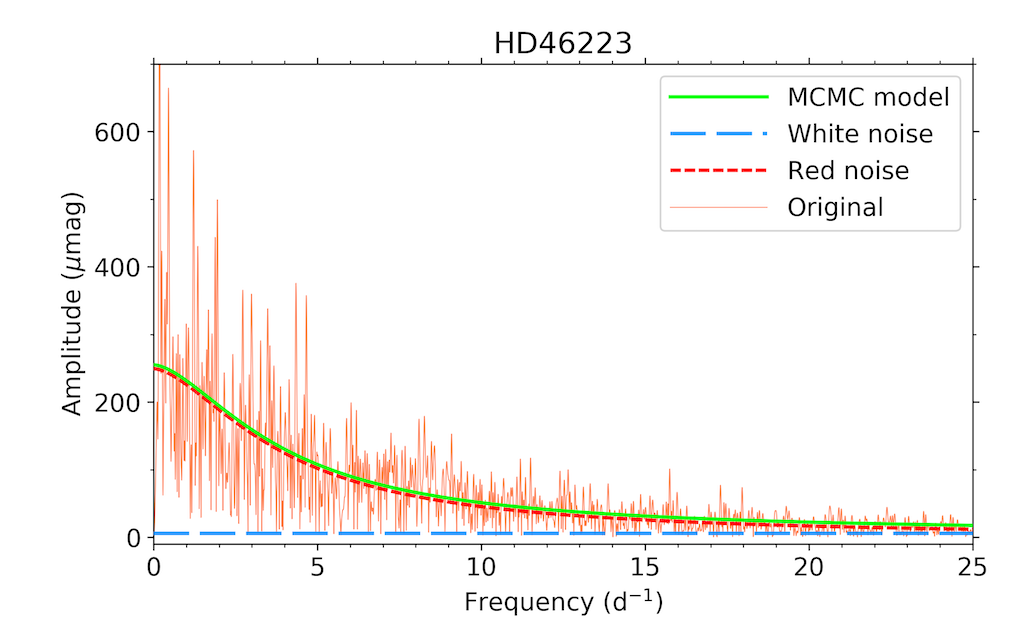}
	\includegraphics[width=0.99\columnwidth]{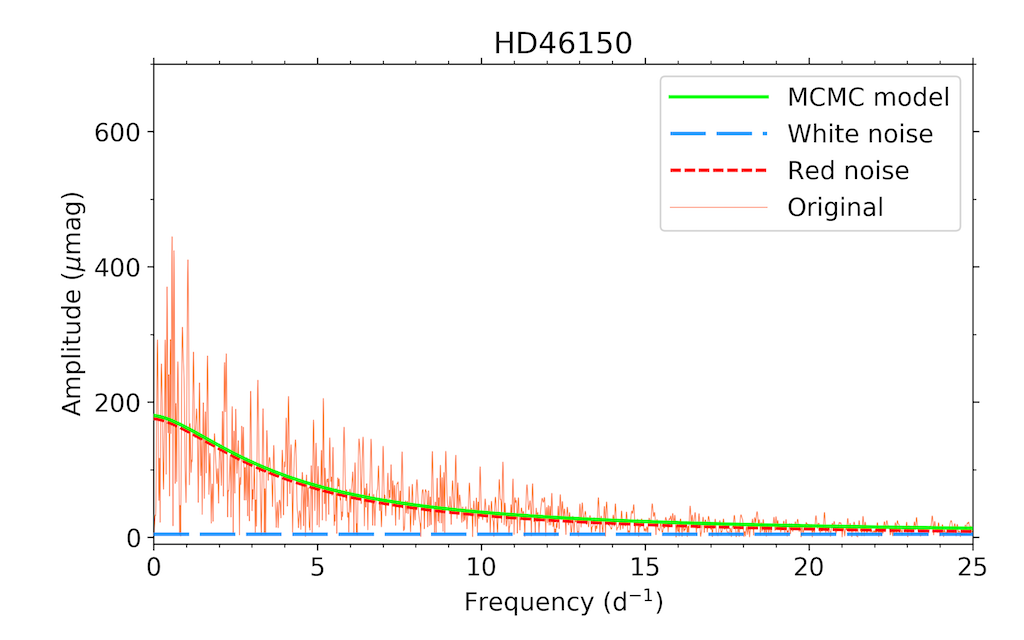}
	\includegraphics[width=0.99\columnwidth]{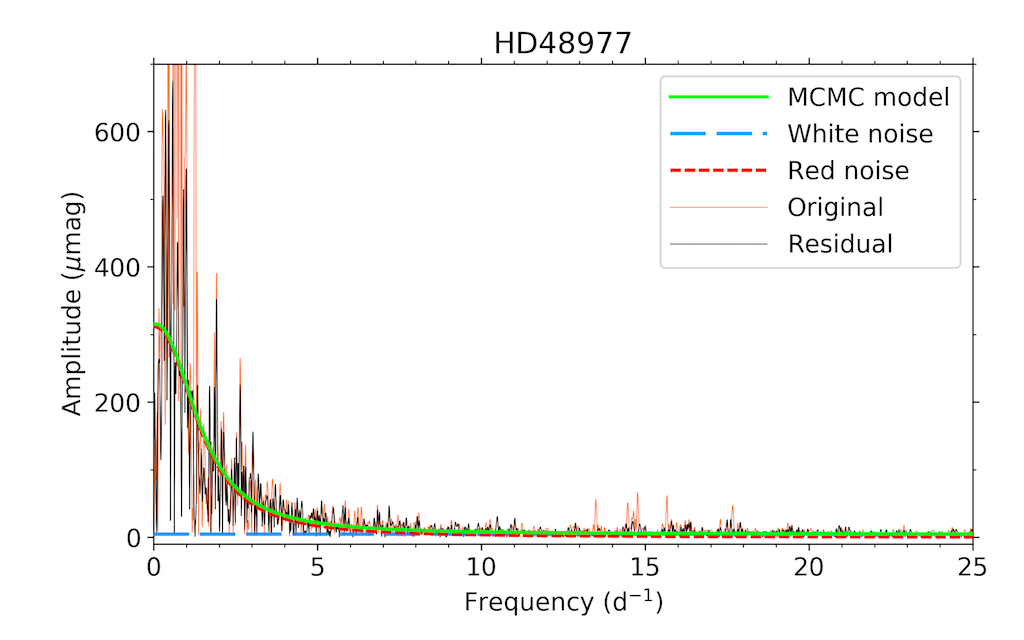}
	\caption{Fitted amplitude spectra calculated using TESS light curves for the O4\,V((f)) star HD~46223 (top panel), the O5\,V((f)) star HD~46150 (middle panel), the B2.5\,V star HD~48977 (bottom panel), which were previously observed by the CoRoT mission and concluded to exhibit IGWs \citep{Bowman2019a}. We note that of these three examples, only HD~48977 underwent iterative pre-whitening to produce a residual amplitude spectrum because it exhibits significant p-mode frequencies above 10~d$^{-1}$.}
	\label{figure: TESS}
	\end{figure}

	We provide three examples of massive stars with fitted stochastic low-frequency variability in Fig.~\ref{figure: TESS} that were previously observed by CoRoT. These stars demonstrate that the broad frequency excess occurs in both the TESS data of massive stars but also in the completely independent CoRoT observations from approximately a decade ago. In Fig.~\ref{figure: TESS}, orange lines denote the original TESS amplitude spectra (before iterative pre-whitening), and black lines denote residual amplitude spectra after iterative pre-whitening has removed S/N~$\geq 4$ frequencies. In Fig.~\ref{figure: TESS}, the solid green, short-dashed red and long-dashed blue lines denote the best-fitting model (cf. Eq.~(\ref{equation: red noise})) and its red- and white-noise components, respectively. Therefore, TESS observations of massive stars confirm the same stochastic broad-frequency excess as was observed by CoRoT \citep{Blomme2011b, Bowman2019a}. 
	
	It is clear from Fig.~\ref{figure: TESS} that more massive O~dwarf stars (e.g. HD~46223 and HD46150; \citealt{Blomme2011b}) have a broader frequency excess that is significant above the white noise level and reaches much higher frequencies compared to B~dwarf stars (e.g. HD~48977; \citealt{Thoul2013}). The stochastic variability of the O~dwarfs is significant over the 1-sector TESS white noise level of order a few $\mu$mag at frequencies higher than $\sim$30~d$^{-1}$ ($\sim$350~$\mu$Hz). Whereas in the B~dwarfs, the stochastic low-frequency variability in the residual amplitude spectrum after coherent pulsation modes have been removed is significant up to $\sim$10~d$^{-1}$ ($\sim$120~$\mu$Hz). Such a frequency dependence of the stochastic variability on the spectral type is common throughout our sample, as demonstrated by the amplitude spectra provided in Appendix~\ref{section: appendix: figures}.
	
	Moreover, we emphasise the ubiquitous detection of stochastic photometric variability in our sample of 70 O and early-B stars. In more massive O~stars, this is the dominant form of variability. However, in a few early-B stars (e.g. HD~34816 and HD~46328) a series of harmonics are present in their amplitude spectra, which are indicative of rotational modulation and/or binarity. Also, some early-B stars in our sample show clear signatures of low-frequency coherent gravity-mode pulsations (e.g. HD~35912 and HD~57539), or high-frequency pressure-mode pulsations (e.g. HD~37209 and HD~37481). We refer the reader to \citet{Burssens2020a*} for the detailed frequency analysis of the coherent pulsators in the sample.


	\subsection{Asteroseismic HR diagrams}
	\label{subsection: HRD}
	
\begin{figure*}
	\centering
	\includegraphics[width=0.99\textwidth]{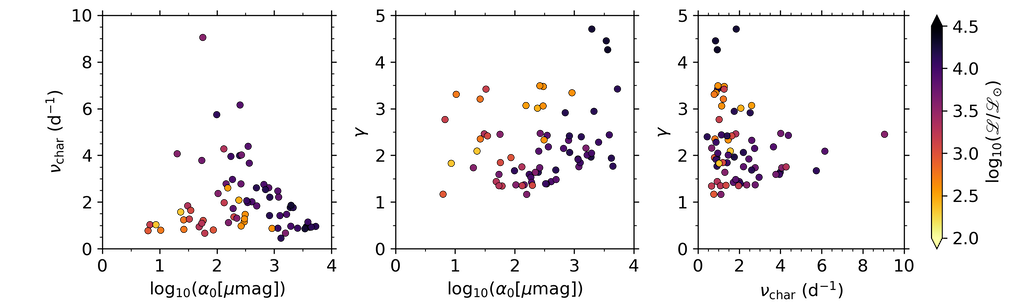}
	\includegraphics[width=0.49\textwidth]{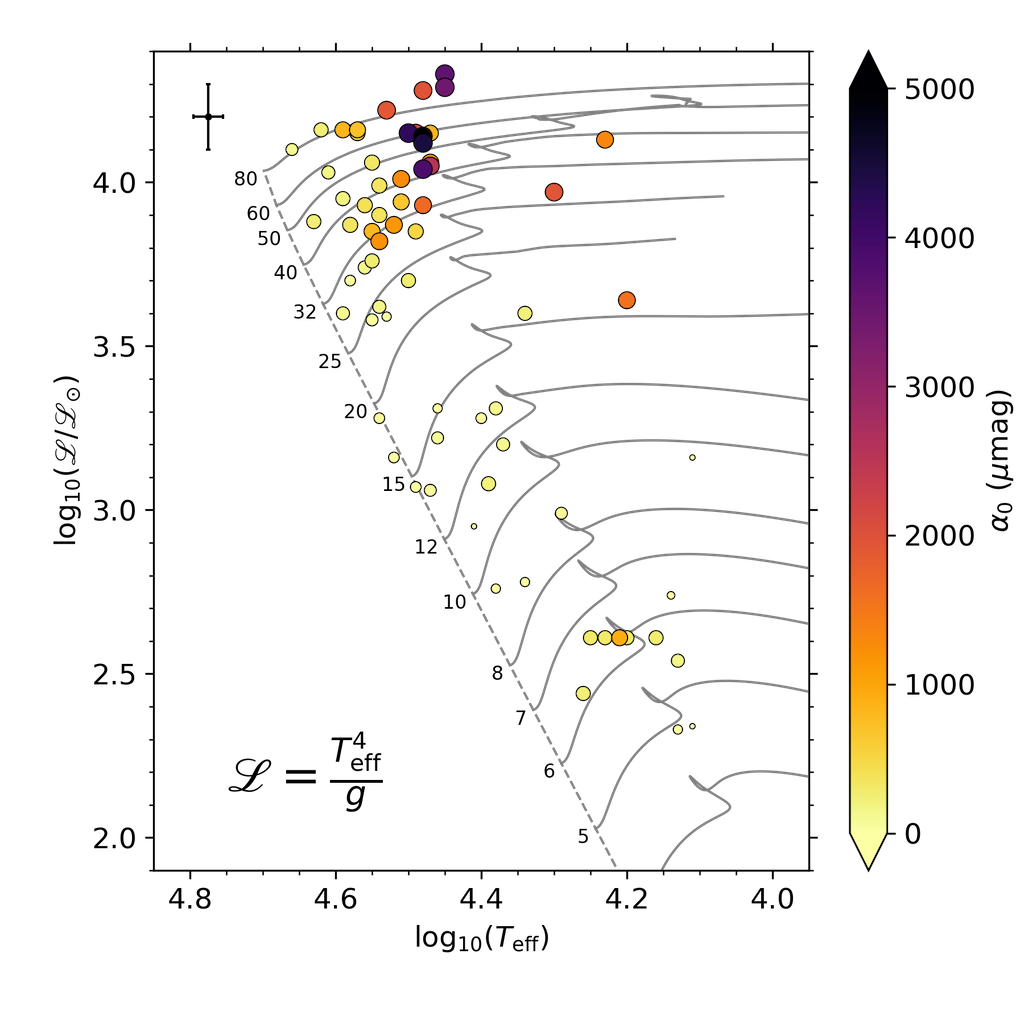}
	\includegraphics[width=0.49\textwidth]{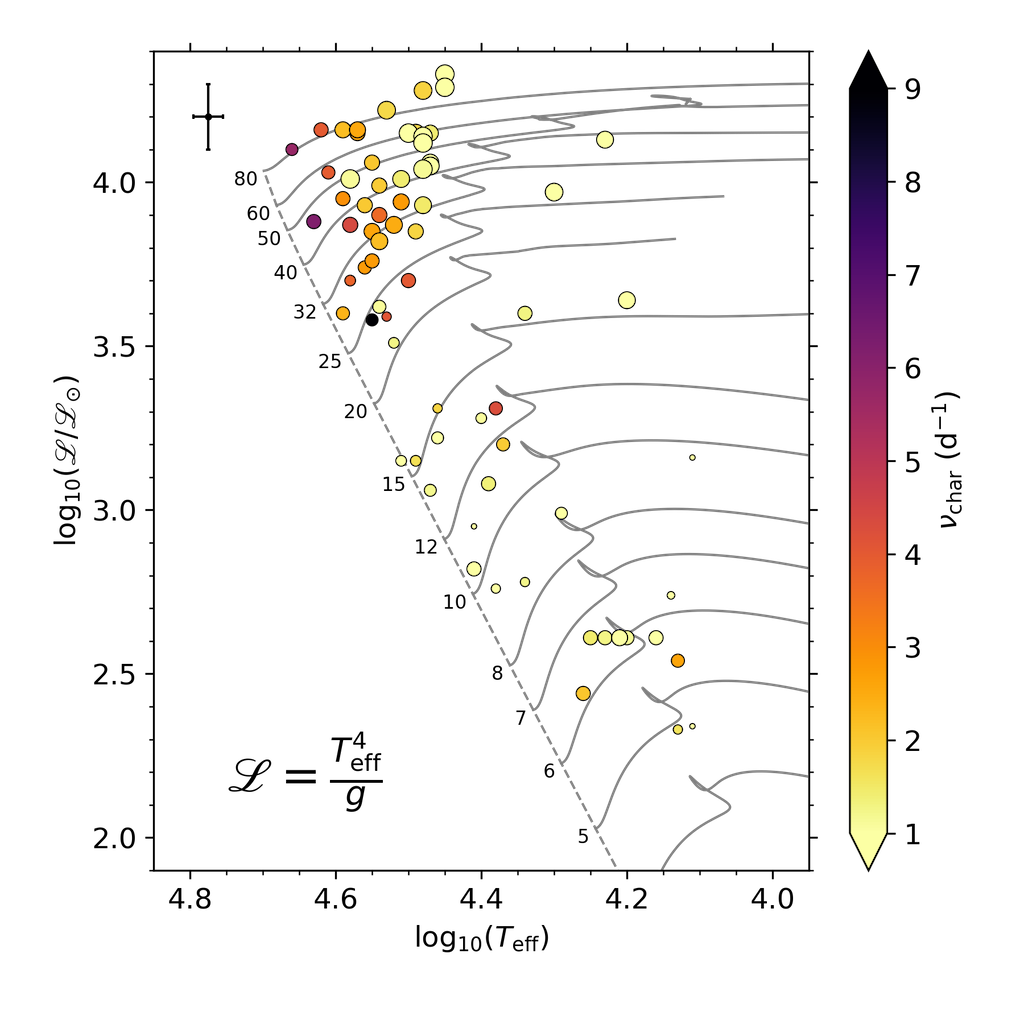}
	\caption{{\it Top row:} pairwise relationship between $\alpha_0$, $\nu_{\rm char}$ and $\gamma$ (cf. Eq.~\ref{equation: red noise}) best-fit parameters for our sample of OB stars. {\it Bottom row:} location of stars in the spectroscopic HR~diagram as filled circles that are colour-coded by the best-fit parameters $\alpha_{0}$ (left) and $\nu_{\rm char}$ (right), and have a symbol size proportional to the fit parameter $\alpha_0$. Evolutionary tracks (in units of M$_{\rm \odot}$) from \citet{Burssens2020a*} are shown as solid grey lines and the dashed grey line represents the ZAMS. A typical spectroscopic error bar for our sample is shown in the top-left corner.}
	\label{figure: HRD}
	\end{figure*}

	To more accurately investigate the observed photometric variability, we place our stars in the spectroscopic HR (sHR) diagram in Fig.~\ref{figure: HRD}. This common approach when studying massive stars involves using the effective stellar luminosity on the ordinate axis defined as: $\mathscr{L} := T_{\rm eff}^{4} / g$ \citep{Langer2014a}. This has the significant advantage of avoiding issues pertaining to uncertain distances and reddening propagating into bolometric luminosity calculations. Each star is represented by a circle in the spectroscopic HR~diagrams in the bottom-left and bottom-right panels of Fig.~\ref{figure: HRD}, which have been colour-coded by the fit parameters $\alpha_{0}$ and $\nu_{\rm char}$, respectively, and have a symbol size that is proportional to the fit parameter $\alpha_{0}$. To illustrate the distribution of our sample in terms of spectroscopic mass and evolutionary stage, we plot the non-rotating evolutionary tracks for initial masses between 4 and 80~M$_{\odot}$ calculated by \citet{Burssens2020a*} as grey lines in Fig.~\ref{figure: HRD}, and a indicative ZAMS line as the dashed-grey line.
	
	In the top row of Fig.~\ref{figure: HRD}, we also plot the pairwise relationship between the individual fit parameters $\alpha_{0}$, $\nu_{\rm char}$ and $\gamma$ as filled circles, which have been colour-coded by the spectroscopic luminosity of each star. For the most massive stars within our sample ($M > 20$~M$_{\odot}$), it is clear that the more luminous stars have larger $\alpha_{0}$ values, such that they have larger amplitudes in their stochastic photometric variability. Furthermore, as best evidenced by the spectroscopic HR~diagrams in the bottom row of Fig.~\ref{figure: HRD}, more massive and more evolved stars not only have larger amplitudes in their stochastic photometric variability, but smaller $\nu_{\rm char}$ values, such that their dominant photometric variability is constrained to longer periods. This is a characteristic signature of IGWs probing stellar evolution: more evolved stars have larger radii, hence IGWs have longer periods. 
	
	Inferring relationships among fit parameters and the distribution in the HR~diagram for stars with masses between approximately $5 \lesssim M \lesssim 20$~M$_{\odot}$ is less clear. This is expected because the stars in this mass regime have diverse causes for their variability, such as rotational modulation and coherent pulsation modes excited by the opacity mechanism ($\beta$~Cephei and Slowly Pulsating B stars; \citealt{ASTERO_BOOK}) superimposed on their stochastic low-frequency variability. In the case of short-length time series and a large number of independent pulsation modes, this makes the extraction of the morphology of the stochastic low-frequency variability subject to somewhat larger uncertainties \citep{Bowman2019a}. Such a photometric result is supported by the diverse range in spectroscopic variability and broadening mechanisms found in main-sequence B~stars (see e.g. \citealt{Simon-Diaz2017a}). 
	
	 In their study of the initial detection of stochastic variability in three O stars observed by CoRoT, which included the O~dwarfs HD~46223 and HD~46150 in this work, \citet{Blomme2011b} discussed an apparent dichotomy in the photometric variability of massive stars. More specifically, early-O stars typically have stochastic low-frequency variability, and late-O and early-B stars typically have coherent pulsation modes. The transition between these takes place at spectral type of around O8, as also illustrated by the star HD~46149 \citep{Degroote2010b}. This spectral type approximately corresponds to a mass of approximately 20~M$_{\odot}$. A similar dichotomy of stars later than O8 having variability caused by coherent pulsation modes has also found in spectroscopic variability studies of early-type stars (e.g. \citealt{Simon-Diaz2017a}). 
	 
	 Our much larger sample of massive stars compared to the previous studies by \citet{Blomme2011b} and \citet{Bowman2019a}, clearly supports the importance of stochastic photometric variability in massive stars. Furthermore, our results are the first evidence that such photometric variability is increasingly important for more massive stars, and is related to the mass and evolutionary stage of the star, as evidenced by Fig.~\ref{figure: HRD}. The parameter space in the spectroscopic HR diagram occupied by the O~stars is also where macroturbulence is the dominant broadening mechanism \citep{Simon-Diaz2017a}. We explore the connection between macroturbulence and stochastic photometric variability in Section~\ref{subsection: macro}.


	\subsection{Waves and macroturbulence}
	\label{subsection: macro}
		
	In addition to the location in the spectroscopic HR~diagram, we are able to probe the relationship between stochastic photometric variability measured in TESS photometry and macroturbulence measured in spectroscopy for a large number of massive stars. Among main-sequence B~stars, it has been noted by several studies that non-radial g-mode pulsations are a plausible physical mechanism to explain macroturbulence given their dominant horizontal velocities \citep{Aerts2009b, Aerts2015c, Simon-Diaz2010b, Simon-Diaz2014a, Simon-Diaz2017a}. Using spectroscopic observations and detailed simulations of broadening and variability in spectral line profiles of main-sequence B stars, \citet{Aerts2009b} demonstrated that macroturbulence is well explained by non-radial coherent g-mode pulsations. \citet{Aerts2015c} extended this study to demonstrate that the collective power of an ensemble of IGWs excited by core convection is also a plausible mechanism for macroturbulent broadening. Gravity waves (coherent and damped) have a ratio in their horizontal to vertical velocities that ranges approximately between 10 and 100 in observations \citep{DeCat2002c, Aerts2003a}, with similar values predicted near the line-forming region by hydrodynamical simulations of IGWs excited by core convection \citep{Rogers2013b, Aerts2015c, Horst2020a*}. 
	
	Given the scarce number of stars more massive than some 20~M$_{\odot}$ with coherent g-mode pulsations, \citet{Grassitelli2015a} demonstrated how the predicted amplitude of turbulent pressure fluctuations originating from the iron subsurface convection zone correlate with observed macroturbulence in massive O~stars. Therefore, IGWs launched by turbulent pressure in the envelopes of massive O~stars has also been inferred to be a plausible mechanism to explain macroturbulence \citep{Grassitelli2015a}. However, surface mechanisms, such as subsurface convection, cause small-scale and time-independent microturbulence in massive stars \citep{Cantiello2009a}, as they cannot reproduce the large-scale required tangential velocity field throughout stellar interiors including the line-forming region \citep{Aerts2009b, Aerts2015c}.
	
	\begin{figure*}
	\centering
	\includegraphics[width=0.99\textwidth]{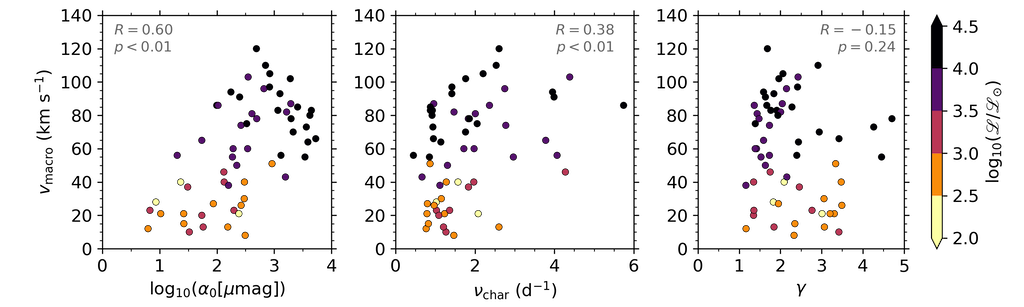}
	\caption{Relationship between $\alpha_0$, $\nu_{\rm char}$ and $\gamma$ (cf. Eq.~\ref{equation: red noise}) and spectroscopic measured of macroturbulence ($v_{\rm macro}$), colour-coded by spectroscopic luminosity. The Spearman's rank correlation coefficient, $R$, and the corresponding $p$-value (obtained from a $t$-test) are also provided.}
	\label{figure: vmacro}
	\end{figure*}
	
	In Fig.~\ref{figure: vmacro} we test the correlation of stochastic photometric variability and macroturbulent broadening measured using high-resolution spectroscopy using the 59 out of 70 stars in our sample with reliable estimates of macroturbulence. We find a clear correlation between the amplitude of the stochastic photometric variability ($\alpha_{0}$) and $v_{\rm macro}$ within our sample. We provide the Spearman’s rank correlation coefficient, $R$, and the corresponding $p$-value (obtained from a $t$-test) in Fig.~\ref{figure: vmacro}. A linear regression reveals a strong correlation ($R = 0.60$; $p < 0.01$) between the photometric amplitudes of stochastic variability in TESS photometry and spectroscopically measured macroturbulence. Our statistical analysis also reveals a significant correlation ($R = 0.38$; $p < 0.01$) between the measured $\nu_{\rm char}$ in the stochastic photometric variability and the spectroscopic macroturbulent broadening. This is expected for IGWs as they are sensitive to the mass and radius of a star. Finally, our results demonstrate that macroturbulence has no significant correlation ($R = -0.15$; $p = 0.24$) with the steepness of the observed frequency spectrum.
	
	The observed relationship between $v_{\rm macro}$ and the fit parameters of the stochastic photometric variability shown in Fig.~\ref{figure: vmacro} builds on the previous theoretical work by \citet{Aerts2015c} and \citet{Grassitelli2015a}, and spectroscopy by \citet{Simon-Diaz2017a}. It demonstrates the importance of IGWs in massive stars. Furthermore, our study provides photometric evidence that IGWs are increasingly more important for stars with larger spectroscopic masses and luminosities. Non-radial pulsations (coherent modes and/or travelling waves --- i.e. IGWs) are unique in their ability to explain the required large-scale tangential velocity field near the stellar surface, given such strong correlations with macroturbulence measured spectroscopically.


\section{Discussion}
\label{section: discussion}

	\subsection{Ubiquitous stochastic variability in photometry}
	\label{subsection: morphology}
	
\begin{figure}
	\centering
	\includegraphics[width=0.49\textwidth]{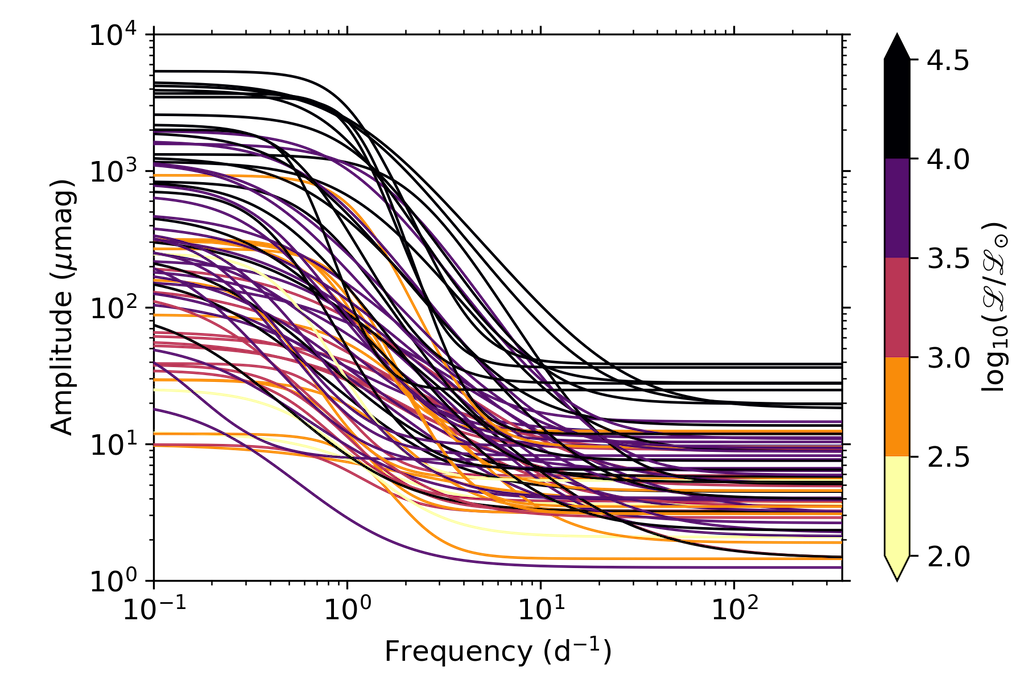}
	\caption{Morphology of the stochastic low-frequency variability (cf. Eq.~\ref{equation: red noise}) determined from TESS light curves for our sample of OB stars, which are colour-coded by spectroscopic luminosity.}
	\label{figure: multi star FT}
	\end{figure}

	In this work, we provide quantitative results for measuring the stochastic low-frequency variability in photometry for the largest number of massive stars to date. As demonstrated by Fig.~\ref{figure: TESS} (and the figures in appendix~\ref{section: appendix: figures}), the frequency range of the measured variability is very broad. To demonstrate the common morphology in the observed stochastic photometric variability in our sample of 70 massive stars, we plot all the fitted profiles using Eq.~(\ref{equation: red noise}) in Fig.~\ref{figure: multi star FT}, which are colour coded by each star's spectroscopic luminosity. Clearly a common morphology exists for all early-type stars, but as mentioned in Section~\ref{subsection: HRD}, the relationship between the observed stochastic photometric variability and the spectroscopic parameters of a star is most evident for the most massive stars in our sample ($M \geq 20$~M$_{\odot}$). The variance within the morphologies for the main-sequence B~stars is quite diverse as expected.
			
	The important features in ascertaining the physical mechanism for the stochastic low-frequency variability in photometry of massive stars are the steepness of the measured amplitude spectrum and the frequency range for the dominant variability \citep{Bowman2019a, Bowman2019b}. Such features are well characterised by the fit parameters $\gamma$ and $\nu_{\rm char}$, respectively. The remaining fit parameters given in Eq.~(\ref{equation: red noise}) are more dependent on the photometric data. To preserve the homogeneity of the TESS photometry, we do not combine it with light curves from different telescopes. For example, the amplitude of the stochastic variability at zero-frequency, $\alpha_{0}$, is a function of the wavelength range of the observations, since the light curves from which amplitude spectra are calculated are not bolometric but in fact wavelength dependent --- specifically ${\rm d}F_{\lambda} / F_{\lambda}$ and not ${\rm d}L/L$. Also, the white noise amplitude, $C_{\rm w}$, is dependent on the number, length and photometric precision of the data points in the light curve.

	
	\subsection{Comparison to previous photometric studies}
	\label{subsection: previous studies}
	
	 Previous studies by \citet{Blomme2011b, Aerts2015c, Bowman2019a} were limited to a few O stars observed by the CoRoT mission. Similarly, \citet{Bowman2019b} studied the stochastic photometric variability in 167 OB stars observed by the K2 and TESS missions. However, parameters from high-resolution spectroscopy were not yet available, and the location of these stars in the colour-magnitude diagram using Gaia photometry \citep{Gaia2016a, Gaia2018a} based on distance \citep{Bailer-Jones2018c}, reddening and extinction estimates \citep{McCall2004b, Green_G_2018a}, meant that masses and evolutionary stages could not be inferred for their sample. Nevertheless, the measured morphology of the stochastic low-frequency variability in this large sample of galactic and extra-galactic OB stars studied by \citet{Bowman2019b} yielded $\gamma \leq 3.5$ for the vast majority of stars. This is fully in agreement with our current study using TESS light curves.
	 
	 Most importantly, the steepness of the amplitude spectrum, $\gamma$, was found to be insensitive to the metallicity of the star, since the sample of O and B stars studied by \citet{Bowman2019b} included 114 ecliptic stars (i.e. $Z \geq Z_{\odot}$) observed by the K2 mission and 53 metal-poor stars (i.e. $Z \simeq 0.5\,Z_{\odot}$) within the Large Magellanic Cloud (LMC) galaxy. The properties of the predicted variability caused by (sub)surface convection are determined by the efficiency of convection in the iron opacity bump, and thus the metallicity of a star \citep{Cantiello2009a, Grassitelli2015a, Lecoanet2019a}, but also the presence of a magnetic field (see e.g. \citealt{Sundqvist2013b}). A study of how magnetic fields systematically affect the presence of stochastic photometric variability and macroturbulence in massive stars requires detections of these two phenomena for a large sample of magnetic stars, which are currently not available.

	  However, in the case of IGWs excited by core convection, only the radius and (convective core) mass of the star set the dominant frequency range ($\nu \lesssim \nu_{\rm char}$) of the IGW amplitude spectrum with a similar steepness ($\gamma$). Asteroseismology of coherent g-mode pulsations has recently allowed the convective core masses of 24 Slowly Pulsating B stars observed by the Kepler mission, which cover the mass range $[3,9]\,$M$_\odot$, to be determined \citep{Pedersen_PhD}. The simultaneous detection of IGWs in many of these main sequence B~stars also allows the efficiency of driving waves by core convection to be tested, but this is beyond the scope of the current work.
	 
	 Our TESS sample comprises a large number of OB stars with masses above 5~M$_{\odot}$ across the southern ecliptic hemisphere observed by TESS. The measured values of the steepness ($\gamma$) and dominant frequency range ($\nu \lesssim \nu_{\rm char}$) of stochastic low-frequency photometric variability are in full agreement with previous observational findings based on massive stars observed by the CoRoT and K2 space missions \citep{Bowman2019a, Bowman2019b}. Therefore, our new TESS results provide further observational evidence that stochastic variability in massive stars is caused by IGWs, either from turbulent core convection and/or from the turbulent pressure fluctuations in subsurface convection zones in their outer envelopes. However, no variability mechanism other than IGWs excited by turbulent convection in the deep interior of stars (i.e. from the convective core during the main sequence and/or shell-burning for post-main sequence stars) is able to explain the similar morphology that extends to relatively large frequencies, the similar $\gamma$ values, and the large $\nu_{\rm char}$ values for metal-poor and metal-rich stars, across such a wide range of masses and evolutionary stages of stars.

	
	\subsection{Comparison to 2D and 3D simulations}
	\label{subsection: simulations}
	
	Recent 3D numerical simulations using a physical stellar structure model as the input reference state predict that core convection produces an IGW amplitude spectrum compatible with a frequency exponent of $0.8 \leq \gamma \leq 3$ \citep{Edelmann2019a}. Conversely, numerical simulations using a Cartesian box geometry predict much steeper frequency spectra \citep{Couston2018b}. The exact properties of the resultant IGW spectrum near the surface of a star depend on its mass, radius and interior rotation rate, which have yet to be fully explored using a parameter study based on hydrodynamical simulations. Nonetheless, the agreement between predicted $\gamma$ values for IGWs excited by core convection from current 3D hydrodynamical simulations and observed $\gamma$ values for our sample of OB stars is striking. 
	
	\begin{figure}
	\centering
	\includegraphics[width=0.49\textwidth]{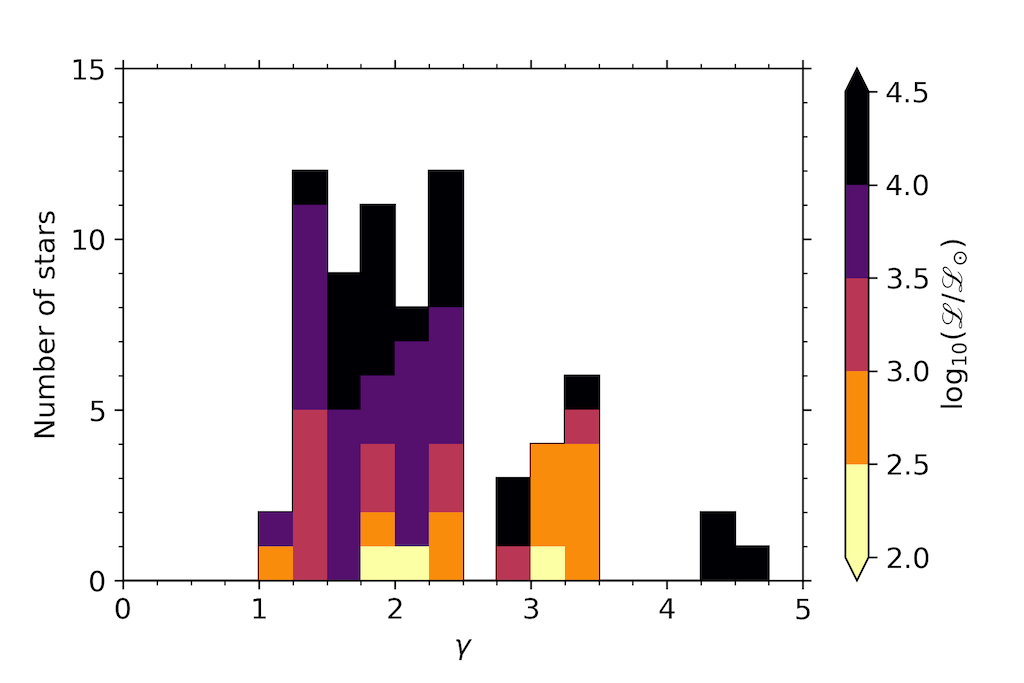}
	\caption{Histogram of the steepness parameter $\gamma$ (cf. Eq.~\ref{equation: red noise}) of the stochastic low-frequency variability for our sample of OB stars colour-coded by spectroscopic luminosity.}
	\label{figure: gamma histogram}
	\end{figure}
	
	We plot the histogram of the measured $\gamma$ values for our 70 stars in Fig.~\ref{figure: gamma histogram}, which is colour-coded using each star’s spectroscopic luminosity similarly to \citet{Bowman2019a}. Our analysis reveals that all massive stars observed by TESS have $\gamma < 5$ with the vast majority having $1 \leq \gamma \leq 3.5$. Furthermore, the observed stochastic variability is significant up to a relatively high frequency regime of tens of d$^{-1}$ in many of the stars. Such a broad frequency range can be explained by an entire spectrum of IGWs, which includes a large range of spatial scales \citep{Edelmann2019a, Horst2020a*}.


\section{Conclusions}
\label{section: conclusions}

In this work we have assembled a sample of 70 massive stars that have spectroscopic parameters determined from high-resolution ground-based spectroscopy by the IACOB project (\citealt{Simon-Diaz2017a, Holgado2018a}, and references therein), and high-precision time-series photometry from the TESS mission \citep{Burssens2020a*}. We applied the methodology devised by \citet{Bowman2019a} and further developed by \citet{Bowman2019b} to our sample of stars to measure the morphological properties of their stochastic low-frequency variability. This first involved removing any significant frequencies associated with rotational modulation and/or coherent pulsation modes via iterative pre-whitening, and subsequently fitting the residual amplitude spectra using a Bayesian MCMC framework to determine the amplitude ($\alpha_{0}$), dominant frequency range ($\nu \lesssim \nu_{\rm char}$), steepness ($\gamma$), and white-noise term ($C_{\rm w}$). Our sample of stars and their determined fit parameters are provided in Tables~\ref{table: stars} and \ref{table: params}, respectively.

We place our sample in the spectroscopic HR~diagram using the accurate parameters from ground-based spectroscopy and demonstrate that the morphology of the amplitude spectrum of stochastic photometric variability is sensitive to the spectroscopic luminosity and evolutionary stage of a star, as shown in Fig.~\ref{figure: HRD}. We demonstrate that stochastic photometric variability is increasingly more important in more massive stars, and that the morphology of the variability directly probes the properties of a star. We also find a clear correlation among the amplitude and characteristic frequency of the stochastic photometric variability and the measured macroturbulent broadening in our sample of massive stars. Macroturbulence and spectral line profile variability have been previously associated with non-radial g-mode pulsations for main-sequence B~stars (e.g. \citealt{Aerts2009b, Simon-Diaz2010b, Aerts2015c}) and turbulent pressure fluctuations exciting IGWs for O stars \citep{Grassitelli2015a}. Here we show that the photometric amplitudes of the stochastic variability strongly correlate with the spectroscopic macroturbulence, as shown in Fig.~\ref{figure: vmacro}. Thus, we conclude that our observational study supports the predictions from theoretical and numerical work that IGWs, excited by core convection and/or turbulent pressure fluctuations, are indeed a plausible mechanism for macroturbulent broadening in massive stars \citep{Aerts2015c, Grassitelli2015a, Simon-Diaz2017a}.

Moreover, we find that the measured fit parameters $\nu_{\rm char}$ and $\gamma$ agree with predictions of the amplitude spectrum of IGWs excited by core convection in hydrodynamical simulations \citep{Rogers2013b, Edelmann2019a, Horst2020a*}. The distribution in the steepness of the observed amplitude spectra, $\gamma$, is shown in Fig.~\ref{figure: gamma histogram}. Clearly, the excitation, propagation, and detectability of IGWs is an important issue for massive stars from theoretical, hydrodynamical, and observational perspectives \citep{Lecoanet2013a, Shiode2013, Rogers2013b, Aerts2015c, Aerts2018a, Aerts2019a, Grassitelli2015a, Augustson2019a, Bowman2019a, Bowman2019b, Edelmann2019a, Horst2020a*}. Our results demonstrate a requirement to include the mixing and angular momentum transport caused by IGWs in the next generation of stellar structure and evolution models, especially for massive stars on the main sequence. In turn this may alleviate the large discrepancies between predicted interior rotation rates from current angular momentum transport theory and observations for stars born with a convective core \citep{Aerts2019b}.

Finally, our results are useful for guiding future asteroseismic studies of massive stars, such as the most massive O~stars and blue supergiants (e.g. \citealt{Saio2006b, Kraus_M_2015b, Bowman2019b}), which may not be pulsating in coherent p- and/or g-modes but do exhibit photometric variability because of IGWs. In the future, we will expand our study to include all massive stars in the IACOB database, with long-term light curves assembled as part of the nominal and extended TESS mission. To more accurately explore the parameter space beyond the main sequence, we will also extend our methodology to include blue supergiants in both the Galaxy and the LMC, such that we can investigate the role of metallicity on the driving mechanism(s) of IGWs in massive stars.


\begin{acknowledgements}
The authors thank the TESS science team for the excellent data and the referee for the supportive and constructive comments. The TESS data presented in this paper were obtained from the Mikulski Archive for Space Telescopes (MAST) at the Space Telescope Science Institute (STScI), which is operated by the Association of Universities for Research in Astronomy, Inc., under NASA contract NAS5-26555. Support to MAST for these data is provided by the NASA Office of Space Science via grant NAG5-7584 and by other grants and contracts. Funding for the TESS mission is provided by the NASA Explorer Program. This research has made use of the SIMBAD database, operated at CDS, Strasbourg, France; the SAO/NASA Astrophysics Data System; and the VizieR catalog access tool, CDS, Strasbourg, France. 

Some of the observations used in this work were obtained with the HERMES spectrograph attached to the Mercator Telescope, operated on the island of La Palma by the Flemish Community, at the Spanish Observatorio del Roque de los Muchachos of the Instituto de Astrof{\' i}sica de Canarias, and further observations obtained with the FEROS spectrograph attached to the 2.2-m MPG/ESO telescope at the La Silla observatory. The HERMES spectrograph is supported by the Fund for Scientific Research of Flanders (FWO), Belgium, the Research Council of KU Leuven, Belgium, the Fonds National Recherches Scientific (FNRS), Belgium, the Royal Observatory of Belgium, the Observatoire de Gen{\' e}ve, Switzerland, and the Th{\" u}ringer Landessternwarte Tautenburg, Germany.

The research leading to these results has received funding from the European Research Council (ERC) under the European Union's Horizon 2020 research and innovation programme (grant agreement No. 670519: MAMSIE). SS-D acknowledges support from the Spanish Government Ministerio de Ciencia e Innovaci\'on through grants PGC-2018-091\,3741-B-C22 and SEV 2015-0548, and from the Canarian Agency for Research, Innovation and Information Society (ACIISI), of the Canary Islands Government, and the European Regional Development Fund (ERDF), under grant with reference ProID2017010115. The work of PVFE was supported by the US Department of Energy through the Los Alamos National Laboratory. Los Alamos National Laboratory is operated by Triad National Security, LLC, for the National Nuclear Security Administration of U.S. Department of Energy (Contract No. 89233218CNA000001). Support for this research was provided by STFC grant ST/S000542/1 and NASA grant NNX17AB92G. The work of LH and FKR is supported by the Klaus Tschira Foundation.

\end{acknowledgements}


\bibliographystyle{aa}
\bibliography{/Users/dominic/Documents/RESEARCH/Bibliography/master_bib}


\begin{appendix}

\section{Extended data tables}
\label{section: appendix: tables}

\onecolumn

\longtab{
\begin{longtable}{l r r r r r r}
\caption{\label{table: stars} Parameters of OB stars studied in this work including common name, TIC number, spectral type, effective temperature ($\log_{10}({\rm T}_{\rm eff})$), spectroscopic luminosity ($\log_{10}(\mathscr{L}/\mathscr{L_{\odot}})$), where $\mathscr{L} := T_{\rm eff}^{4} / g$, projected surface rotational velocity ($v\,\sin\,i$), and macroturbulent broadening ($v_{\rm macro}$), which are taken from \citet{Burssens2020a*}. The typical uncertainty for $\log_{10}({\rm T}_{\rm eff})$ is estimated to be 0.03~dex, and for $\log_{10}(\mathscr{L}/\mathscr{L_{\odot}})$ estimated to be 0.15~dex (see \citealt{Simon-Diaz2017a, Holgado2018a}).} \\
\hline\hline
\multicolumn{1}{c}{Name} & \multicolumn{1}{c}{TIC} & \multicolumn{1}{c}{Sp. Type} & \multicolumn{1}{c}{$\log_{10}({\rm T}_{\rm eff})$} & \multicolumn{1}{c}{$\log_{10}(\mathscr{L}/\mathscr{L_{\odot}})$} & \multicolumn{1}{c}{$v\,\sin\,i$} & \multicolumn{1}{c}{$v_{\rm macro}$} \\
\multicolumn{1}{c}{} & \multicolumn{1}{c}{} & \multicolumn{1}{c}{} & \multicolumn{1}{c}{} & \multicolumn{1}{c}{} & \multicolumn{1}{c}{(km\,s$^{-1}$)} & \multicolumn{1}{c}{(km\,s$^{-1}$)} \\
\hline
\endfirsthead
\caption{\it continued.}\\
\hline\hline
\multicolumn{1}{c}{Name} & \multicolumn{1}{c}{TIC} & \multicolumn{1}{c}{Sp. Type} & \multicolumn{1}{c}{$\log_{10}({\rm T}_{\rm eff})$} & \multicolumn{1}{c}{$\log_{10}(\mathscr{L}/\mathscr{L_{\odot}})$} & \multicolumn{1}{c}{$v\,\sin\,i$} & \multicolumn{1}{c}{$v_{\rm macro}$} \\
\multicolumn{1}{c}{} & \multicolumn{1}{c}{} & \multicolumn{1}{c}{} & \multicolumn{1}{c}{} & \multicolumn{1}{c}{} & \multicolumn{1}{c}{(km\,s$^{-1}$)} & \multicolumn{1}{c}{(km\,s$^{-1}$)} \\
\hline
\endhead
\hline
\endfoot
\multicolumn{7}{l}{\bf O dwarf stars:} \\ 
HD~96715	&	306491594	&	O4\,V((f))z			&	4.66	&	4.10	&	59	&	86	\\
HD~46223	&	234881667	&	O4\,V((f))			&	4.62	&	4.16	&	60	&	91	\\
HD~155913	&	216662610	&	O4.5\,Vn((f))		&	4.63	&	3.88	&	278	&	$-$	\\
HD~46150	&	234840662	&	O5\,V((f))			&	4.61	&	4.03	&	71	&	94	\\
HD~90273	&	464295672	&	ON7\,V			&	4.59	&	3.95	&	55	&	55	\\
HD~110360	&	433738620	&	ON7\,V			&	4.59	&	3.60	&	96	&	86	\\
HD~47839	&	220322383	&	O7\,V			&	4.58	&	3.70	&	43	&	65	\\
HD~53975	&	148506724	&	O7.5\,Vz			&	4.56	&	3.74	&	181	&	$-$	\\
HD~41997	&	294114621	&	O7.5\,Vn((f))		&	4.55	&	3.85	&	262	&	$-$	\\
HD~46573	&	234947719	&	O7\,V((f))z			&	4.56	&	3.93	&	77	&	81	\\
HD~48279	&	234009943	&	O8\,V			&	4.55	&	3.76	&	131	&	74	\\
HD~46056	&	234834992	&	O8\,Vn			&	4.55	&	3.58	&	370	&	$-$	\\ 
HD~38666	&	100589904	&	O9.5\,V			&	4.53	&	3.59	&	111	&	56	\\
\hline 
\multicolumn{7}{l}{\bf O subgiant stars:} \\
HD~74920	&	430625455	&	O7.5\,IVn((f))		&	4.54	&	3.90	&	291	&	$-$	\\
HD~135591	&	455675248	&	O8\,IV((f))			&	4.54	&	3.99	&	60	&	60	\\
HD~326331	&	339568114	&	O8\,IVn((f))		&	4.54	&	3.82	&	332	&	$-$	\\
HD~37041	&	427395049	&	O9.5\,IVp			&	4.54	&	3.28	&	134	&	$-$	\\
HD~123056	&	330281456	&	O9.5\,IV(n)		&	4.50	&	3.70	&	193	&	$-$	\\
\hline 
\multicolumn{7}{l}{\bf O giant stars:} \\
HD~97253	&	467065657	&	O5\,III(f)			&	4.59	&	4.16	&	70	&	105	\\
HD~93843	&	465012898	&	O5\,III(fc)			&	4.57	&	4.15	&	58	&	120	\\
HD~156738	&	195288472	&	O6.5\,III(f)			&	4.58	&	3.87	&	65	&	103	\\
HD~36861	&	436103278	&	O8\,III((f))			&	4.55	&	4.06	&	53	&	75	\\
HD~150574	&	234648113	&	ON9\,III(n)			&	4.52	&	3.87	&	252	&	$-$	\\
HD~152247	&	339570292	&	O9.2\,III			&	4.51	&	3.94	&	82	&	96	\\
HD~55879	&	178489528	&	O9.7\,III			&	4.49	&	3.85	&	26	&	60	\\
HD~154643	&	43284243		&	O9.7\,III			&	4.49	&	3.85	&	101	&	78	\\
\hline 
\multicolumn{7}{l}{\bf O bright giant and supergiant stars:} \\
CPD-47~2963	&	30653985		&	O5\,Ifc			&	4.57	&	4.16	&	67	&	110	\\
HD~156154	&	152659955	&	O7.5\,Ib(f)			&	4.53	&	4.22	&	62	&	102	\\
HD~112244	&	406050497	&	O8.5\,Iab(f)p		&	4.50	&	4.15	&	124	&	80	\\
HD~151804	&	337793038	&	O8\,Iaf			&	4.45	&	4.33	&	72	&	73	\\
HD~303492	&	459532732	&	O8.5\,Iaf			&	4.45	&	4.29	&	87	&	55	\\
HD~57061	&	106347931	&	O9\,II			&	4.51	&	4.01	&	57	&	93	\\
HD~152249	&	339567904	&	OC9\,Iab			&	4.49	&	4.15	&	71	&	70	\\
HD~152424	&	247267245	&	OC9.2\,Ia			&	4.48	&	4.14	&	59	&	66	\\
HD~154368	&	41792209		&	O9.5\,Iab			&	4.48	&	4.28	&	65	&	78	\\
HD~152003	&	338640317	&	O9.7\,Iab\,Nwk		&	4.48	&	4.12	&	65	&	83	\\
HD~152147	&	246953610	&	O9.7\,Ib\,Nwk		&	4.48	&	4.04	&	91	&	64	\\
\hline
\multicolumn{7}{l}{\bf B dwarf stars:} \\
HD~36960	&	427373484	&	B0.5\,V			&	4.46	&	3.31	&	23	&	37	\\
HD~37042	&	427395058	&	B0.7\,V			&	4.47	&	3.06	&	33	&	13	\\
HD~43112	&	434384707	&	B1\,V			&	4.41	&	2.95	&	7	&	12	\\
HD~35912	&	464839773	&	B2\,V			&	4.26	&	2.44	&	11	&	21	\\
HD~48977	&	202148345	&	B2.5\,V			&	4.25	&	2.61	&	26	&	8	\\
\hline
\multicolumn{7}{l}{\bf B subgiant stars:} \\
HD~34816	&	442871031	&	B0.5\,IV			&	4.46	&	3.22	&	25	&	$-$	\\
HD~46328	&	47763235		&	B0.5\,IV			&	4.40	&	3.28	&	7	&	20	\\
HD~50707	&	78897024		&	B1\,IV			&	4.38	&	3.31	&	29	&	46	\\
HD~37481	&	332913301	&	B1.5\,IV			&	4.34	&	2.78	&	74	&	21	\\
HD~37209	&	388935529	&	B2\,IV			&	4.38	&	2.76	&	50	&	15	\\
HD~26912	&	283793973	&	B3\,IV			&	4.20	&	2.61	&	53	&	30	\\
HD~37711	&	59215060		&	B3\,IV			&	4.21	&	2.61	&	68	&	51	\\
HD~57539	&	10176636		&	B3\,IV			&	4.13	&	2.54	&	162	&	13	\\
HD~41753	&	151464886	&	B3\,IV			&	4.23	&	2.61	&	25	&	40	\\
HD~224990	&	313934087	&	B5\,IV			&	4.13	&	2.33	&	20	&	40	\\
\hline
\multicolumn{7}{l}{\bf B giant stars:} \\
HD~48434	&	234052684	&	B0\,III			&	4.48	&	3.93	&	48	&	82	\\
HD~61068	&	349043273	&	B2\,III			&	4.39	&	3.08	&	12	&	23	\\
HD~35468	&	365572007	&	B2\,III			&	4.29	&	2.99	&	53	&	27	\\
\hline
\multicolumn{7}{l}{\bf B bright giant and supergiant stars:} \\
HD~44743	&	34590771		&	B1\,II-III			&	4.37	&	3.20	&	24	&	40	\\
HD~54764	&	95513457		&	B1\,II				&	4.30	&	3.97	&	123	&	87	\\
HD~52089	&	63198307		&	B2\,II				&	4.34	&	3.60	&	26	&	50	\\
HD~51309	&	146908355	&	B3\,II				&	4.20	&	3.64	&	27	&	43	\\
HD~46769	&	281148636	&	B5\,II				&	4.11	&	3.16	&	70	&	23	\\
HD~27563	&	37777866		&	B7\,II				&	4.16	&	2.61	&	34	&	26	\\
HD~53244	&	148109427	&	B8\,II				&	4.14	&	2.74	&	36	&	21	\\
HD~37128	&	427451176	&	B0\,Ia			&	4.47	&	4.05	&	55	&	85	\\
HD~38771	&	66651575		&	B0.5\,Ia			&	4.47	&	4.06	&	53	&	83	\\
HD~53138	&	80466973		&	B3\,Iab			&	4.23	&	4.13	&	37	&	56	\\
HD~39985	&	102281507	&	B9\,Ib			&	4.11	&	2.34	&	26	&	28	\\
\hline
\multicolumn{7}{l}{\bf Peculiar stars:} \\
HD~37061	&	427393920	&	O9.5\,V			&	4.49	&	3.07 &	210	&	$-$	\\
HD~37742	&	11360636		&	O9.2\,Ib\,var\,Nwk	&	4.47	&	4.15	&	122	&	97	\\
HD~57682	&	187458882	&	O9.2\,IV			&	4.54	&	3.62	&	12	&	38	\\
HD~54879	&	177860391	&	O9.7\,V			&	4.52	&	3.16	&	7	&	10	\\
\end{longtable}
}

\longtab{
\begin{longtable}{l r r r r r}
\caption{\label{table: params} Optimised parameters for the morphology of low-frequency variability (cf. Eq.~(\ref{equation: red noise})) using a Bayesian MCMC fitting method.} \\
\hline\hline
\multicolumn{1}{c}{Name} & \multicolumn{1}{c}{TIC} & \multicolumn{1}{c}{$\alpha_0$} & \multicolumn{1}{c}{$\nu_{\rm char}$} & \multicolumn{1}{c}{$\gamma$} & \multicolumn{1}{c}{$C_{\rm W}$} \\
\multicolumn{1}{c}{} & \multicolumn{1}{c}{} & \multicolumn{1}{c}{($\mu$mag)} & \multicolumn{1}{c}{(d$^{-1}$)} & \multicolumn{1}{c}{} & \multicolumn{1}{c}{($\mu$mag)}  \\
\hline
\endfirsthead
\caption{\it continued.}\\
\hline\hline
\multicolumn{1}{c}{Name} & \multicolumn{1}{c}{TIC} & \multicolumn{1}{c}{$\alpha_0$} & \multicolumn{1}{c}{$\nu_{\rm char}$} & \multicolumn{1}{c}{$\gamma$} & \multicolumn{1}{c}{$C_{\rm W}$} \\
\multicolumn{1}{c}{} & \multicolumn{1}{c}{} & \multicolumn{1}{c}{($\mu$mag)} & \multicolumn{1}{c}{(d$^{-1}$)} & \multicolumn{1}{c}{} & \multicolumn{1}{c}{($\mu$mag)} \\
\hline
\endhead
\hline
\endfoot
\multicolumn{6}{l}{\bf O dwarf stars:} \\
HD~96715	&	306491594	&	$99.219	\pm	0.039$	&	$5.74308	\pm	0.00368$	&	$1.66987	\pm	0.00101$	&	$3.210	\pm	0.007$	\\
HD~46223	&	234881667	&	$249.330	\pm	0.047$	&	$3.99087	\pm	0.00135$	&	$1.63706	\pm	0.00041$	&	$5.766	\pm	0.002$	\\
HD~155913	&	216662610	&	$253.521	\pm	0.033$	&	$6.15743	\pm	0.00115$	&	$2.08883	\pm	0.00054$	&	$6.635	\pm	0.006$	\\
HD~46150	&	234840662	&	$175.470	\pm	0.044$	&	$3.95351	\pm	0.00181$	&	$1.59155	\pm	0.00050$	&	$4.480	\pm	0.002$	\\
HD~90273	&	464295672	&	$186.796	\pm	0.051$	&	$2.96658	\pm	0.00147$	&	$1.52985	\pm	0.00051$	&	$6.361	\pm	0.002$	\\
HD~110360	&	433738620	&	$104.664	\pm	0.061$	&	$2.36468	\pm	0.00242$	&	$1.36668	\pm	0.00116$	&	$11.971	\pm	0.007$	\\
HD~47839	&	220322383	&	$54.461	\pm	0.046$	&	$3.78150	\pm	0.00548$	&	$1.59141	\pm	0.00212$	&	$3.900	\pm	0.006$	\\
HD~53975	&	148506724	&	$144.615	\pm	0.055$	&	$2.77860	\pm	0.00194$	&	$1.36739	\pm	0.00071$	&	$2.631	\pm	0.007$	\\
HD~41997	&	294114621	&	$803.647	\pm	0.040$	&	$2.61022	\pm	0.00018$	&	$2.32176	\pm	0.00023$	&	$9.622	\pm	0.004$	\\
HD~46573	&	234947719	&	$409.204	\pm	0.058$	&	$2.00862	\pm	0.00051$	&	$1.43600	\pm	0.00027$	&	$3.883	\pm	0.006$	\\
HD~48279	&	234009943	&	$262.925	\pm	0.052$	&	$2.77586	\pm	0.00089$	&	$1.73500	\pm	0.00053$	&	$11.124	\pm	0.005$	\\
HD~46056	&	234834992	&	$56.700	\pm	0.026$	&	$9.05220	\pm	0.00551$	&	$2.45063	\pm	0.00266$	&	$7.696	\pm	0.006$	\\
HD~38666	&	100589904	&	$20.239	\pm	0.048$	&	$4.06758	\pm	0.01641$	&	$1.73389	\pm	0.00697$	&	$1.246	\pm	0.006$	\\
\hline
\multicolumn{6}{l}{\bf O subgiant stars:} \\
HD~74920	&	430625455	&	$367.615	\pm	0.036$	&	$3.66392	\pm	0.00053$	&	$2.18152	\pm	0.00040$	&	$4.029	\pm	0.005$	\\
HD~135591	&	455675248	&	$343.595	\pm	0.061$	&	$1.97700	\pm	0.00066$	&	$1.38857	\pm	0.00026$	&	$2.235	\pm	0.002$	\\
HD~326331	&	339568114	&	$1196.544	\pm	0.056$	&	$2.20987	\pm	0.00017$	&	$1.75566	\pm	0.00012$	&	$9.365	\pm	0.005$	\\
HD~37041	&	427395049	&	$48.977	\pm	0.085$	&	$0.93624	\pm	0.00304$	&	$1.43815	\pm	0.00304$	&	$4.910	\pm	0.005$	\\
HD~123056	&	330281456	&	$268.094	\pm	0.033$	&	$4.01672	\pm	0.00065$	&	$2.46953	\pm	0.00073$	&	$8.208	\pm	0.005$	\\
\hline 
\multicolumn{6}{l}{\bf O giant stars:} \\
HD~97253	&	467065657	&	$837.597	\pm	0.048$	&	$2.20426	\pm	0.00019$	&	$2.06179	\pm	0.00020$	&	$5.257	\pm	0.004$	\\
HD~93843	&	465012898	&	$489.559	\pm	0.060$	&	$2.60714	\pm	0.00055$	&	$1.68517	\pm	0.00032$	&	$2.333	\pm	0.006$	\\
HD~156738	&	195288472	&	$350.922	\pm	0.032$	&	$4.38608	\pm	0.00054$	&	$2.42795	\pm	0.00045$	&	$9.671	\pm	0.005$	\\
HD~36861	&	436103278	&	$330.703	\pm	0.064$	&	$2.05434	\pm	0.00077$	&	$1.39102	\pm	0.00030$	&	$1.453	\pm	0.002$	\\
HD~150574	&	234648113	&	$1156.742	\pm	0.049$	&	$2.47023	\pm	0.00017$	&	$1.92152	\pm	0.00014$	&	$14.547	\pm	0.005$	\\
HD~152247	&	339570292	&	$661.325	\pm	0.043$	&	$2.75008	\pm	0.00027$	&	$2.14782	\pm	0.00024$	&	$10.339	\pm	0.004$	\\
HD~55879	&	178489528	&	$191.802	\pm	0.063$	&	$1.71966	\pm	0.00108$	&	$1.42054	\pm	0.00056$	&	$3.195	\pm	0.002$	\\
HD~154643	&	43284243		&	$495.620	\pm	0.061$	&	$1.83226	\pm	0.00041$	&	$1.48131	\pm	0.00021$	&	$5.148	\pm	0.002$	\\
\hline
\multicolumn{6}{l}{\bf O bright giant and supergiant stars:} \\
CPD-47~2963	&	30653985		&	$705.866	\pm	0.043$	&	$2.53356	\pm	0.00019$	&	$2.91261	\pm	0.00047$	&	$6.488	\pm	0.004$	\\
HD~156154	&	152659955	&	$1911.500	\pm	0.056$	&	$1.76469	\pm	0.00008$	&	$1.96280	\pm	0.00009$	&	$13.637	\pm	0.004$	\\
HD~112244	&	406050497	&	$4203.703\pm	0.090$	&	$0.91870	\pm	0.00003$	&	$1.93925	\pm	0.00006$	&	$19.509	\pm	0.004$	\\
HD~151804	&	337793038	&	$3642.777\pm	0.055$	&	$0.93978	\pm	0.00001$	&	$4.26287	\pm	0.00024$	&	$36.297	\pm	0.004$	\\
HD~303492	&	459532732	&	$3447.949\pm	0.057$	&	$0.85247	\pm	0.00001$	&	$4.45373	\pm	0.00027$	&	$11.850	\pm	0.002$	\\
HD~57061	&	106347931	&	$1261.169\pm	0.078$	&	$1.41680	\pm	0.00014$	&	$1.84189	\pm	0.00017$	&	$3.970	\pm	0.005$	\\
HD~152249	&	339567904	&	$2144.886\pm	0.044$	&	$1.76168	\pm	0.00005$	&	$2.94309	\pm	0.00015$	&	$27.729	\pm	0.004$	\\
HD~152424	&	247267245	&	$5315.488\pm	0.059$	&	$0.95329	\pm	0.00001$	&	$3.42169	\pm	0.00011$	&	$38.442	\pm	0.004$	\\
HD~154368	&	41792209		&	$1971.732\pm	0.037$	&	$1.85794	\pm	0.00004$	&	$4.70571	\pm	0.00034$	&	$24.824	\pm	0.004$	\\
HD~152003	&	338640317	&	$4459.742\pm	0.090$	&	$0.92626	\pm	0.00003$	&	$1.76747	\pm	0.00005$	&	$18.221	\pm	0.004$	\\
HD~152147	&	246953610	&	$3879.566\pm	0.060$	&	$1.13960	\pm	0.00002$	&	$2.43834	\pm	0.00008$	&	$27.977	\pm	0.004$	\\

\hline 
\multicolumn{6}{l}{\bf B dwarf stars:} \\
HD~36960	&	427373484	&	$30.868	\pm	0.047$	&	$1.83618	\pm	0.00375$	&	$2.46009	\pm	0.00780$	&	$3.814	\pm	0.004$	\\
HD~37042	&	427395058	&	$57.494	\pm	0.066$	&	$1.20717	\pm	0.00210$	&	$1.84494	\pm	0.00334$	&	$9.116	\pm	0.004$	\\
HD~43112	&	434384707	&	$6.269	\pm	0.117$	&	$0.76982	\pm	0.03045$	&	$1.16679	\pm	0.02216$	&	$3.770	\pm	0.006$	\\
HD~35912	&	464839773	&	$241.602	\pm	0.048$	&	$2.08064	\pm	0.00049$	&	$3.01021	\pm	0.00161$	&	$5.396	\pm	0.005$	\\
HD~48977	&	202148345	&	$311.431	\pm	0.066$	&	$1.46731	\pm	0.00043$	&	$2.33096	\pm	0.00098$	&	$4.565	\pm	0.005$	\\

\hline
\multicolumn{6}{l}{\bf B subgiant stars:} \\
HD~34816	&	442871031	&	$61.383	\pm	0.114$	&	$0.66173	\pm	0.00244$	&	$1.34255	\pm	0.00275$	&	$1.445	\pm	0.005$	\\
HD~46328	&	47763235		&	$54.479	\pm	0.112$	&	$1.08885	\pm	0.00425$	&	$1.34896	\pm	0.00326$	&	$3.490	\pm	0.007$	\\
HD~50707	&	78897024		&	$132.110	\pm	0.035$	&	$4.27661	\pm	0.00197$	&	$1.75206	\pm	0.00061$	&	$2.889	\pm	0.002$	\\
HD~37481	&	332913301	&	$26.284	\pm	0.061$	&	$1.23368	\pm	0.00330$	&	$3.20558	\pm	0.03687$	&	$3.132	\pm	0.005$	\\
HD~37209	&	388935529	&	$26.461	\pm	0.097$	&	$0.82600	\pm	0.00426$	&	$2.35097	\pm	0.02790$	&	$3.195	\pm	0.005$	\\
HD~26912	&	283793973	&	$299.788	\pm	0.061$	&	$1.15152	\pm	0.00028$	&	$3.05793	\pm	0.00195$	&	$3.078	\pm	0.004$	\\
HD~37711	&	59215060		&	$916.636	\pm	0.058$	&	$0.86817	\pm	0.00006$	&	$3.34037	\pm	0.00062$	&	$9.476	\pm	0.004$	\\
HD~57539	&	10176636		&	$154.777	\pm	0.044$	&	$2.60349	\pm	0.00099$	&	$3.06784	\pm	0.00251$	&	$5.565	\pm	0.004$	\\
HD~41753	&	151464886	&	$302.331	\pm	0.048$	&	$1.27728	\pm	0.00023$	&	$3.47522	\pm	0.00175$	&	$12.424	\pm	0.004$	\\
HD~224990	&	313934087	&	$23.296	\pm	0.065$	&	$1.57162	\pm	0.00700$	&	$2.09149	\pm	0.00988$	&	$2.084	\pm	0.004$	\\
\hline 
\multicolumn{6}{l}{\bf B giant stars:} \\
HD~48434	&	234052684	&	$1647.090\pm	0.063$	&	$1.47146	\pm	0.00009$	&	$2.00588	\pm	0.00013$	&	$8.813	\pm	0.004$	\\
HD~61068	&	349043273	&	$198.120	\pm	0.090$	&	$1.36075	\pm	0.00121$	&	$1.35714	\pm	0.00076$	&	$5.256	\pm	0.007$	\\
HD~35468	&	365572007	&	$86.576	\pm	0.078$	&	$0.80155	\pm	0.00110$	&	$1.95155	\pm	0.00305$	&	$1.895	\pm	0.004$	\\
\hline
\multicolumn{6}{l}{\bf B bright giant and supergiant stars:} \\
HD~44743	&	34590771		&	$133.771	\pm	0.080$	&	$1.96975	\pm	0.00220$	&	$1.35040	\pm	0.00106$	&	$8.802	\pm	0.008$	\\
HD~54764	&	95513457		&	$1949.927\pm	0.070$	&	$0.96136	\pm	0.00005$	&	$2.04624	\pm	0.00014$	&	$10.902	\pm	0.004$	\\
HD~52089	&	63198307		&	$222.135	\pm	0.069$	&	$1.30671	\pm	0.00070$	&	$1.63586	\pm	0.00065$	&	$2.109	\pm	0.002$	\\
HD~51309	&	146908355	&	$1572.988\pm	0.087$	&	$0.66617	\pm	0.00006$	&	$2.15494	\pm	0.00023$	&	$5.917	\pm	0.002$	\\
HD~46769	&	281148636	&	$6.763	\pm	0.056$	&	$1.02897	\pm	0.01017$	&	$2.76723	\pm	0.07812$	&	$3.139	\pm	0.004$	\\
HD~27563	&	37777866		&	$264.680	\pm	0.059$	&	$0.97211	\pm	0.00025$	&	$3.49058	\pm	0.00231$	&	$3.479	\pm	0.004$	\\
HD~53244	&	148109427	&	$10.428	\pm	0.065$	&	$0.79102	\pm	0.00547$	&	$3.30717	\pm	0.06935$	&	$1.444	\pm	0.004$	\\
HD~37128	&	427451176	&	$2562.785\pm	0.069$	&	$0.87736	\pm	0.00003$	&	$2.28089	\pm	0.00012$	&	$19.682	\pm	0.004$	\\
HD~38771	&	66651575		&	$1159.940	\pm	0.083$	&	$0.87575	\pm	0.00010$	&	$1.91176	\pm	0.00022$	&	$8.819	\pm	0.004$	\\
HD~53138	&	80466973		&	$1313.016\pm	0.108$	&	$0.44918	\pm	0.00005$	&	$2.39756	\pm	0.00043$	&	$5.089	\pm	0.002$	\\
HD~39985	&	102281507	&	$8.600	\pm	0.081$	&	$1.02655	\pm	0.01553$	&	$1.82852	\pm	0.03437$	&	$3.579	\pm	0.005$	\\

\hline 
\multicolumn{6}{l}{\bf Peculiar stars:} \\
HD~37061	&	427393920	&	$34.886	\pm	0.068$	&	$1.64256	\pm	0.00459$	&	$2.41671	\pm	0.01151$	&	$3.237	\pm	0.005$	\\
HD~37742	&	11360636		&	$829.560	\pm	0.070$	&	$1.41864	\pm	0.00016$	&	$2.41592	\pm	0.00042$	&	$7.544	\pm	0.005$	\\
HD~57682	&	187458882	&	$159.073	\pm	0.095$	&	$1.11609	\pm	0.00138$	&	$1.16316	\pm	0.00080$	&	$3.029	\pm	0.007$	\\
HD~54879	&	177860391	&	$32.960	\pm	0.052$	&	$1.26842	\pm	0.00235$	&	$3.41894	\pm	0.01621$	&	$5.871	\pm	0.004$	\\
\end{longtable}
}


\section{Fitted amplitude spectra figures}
\label{section: appendix: figures}

\begin{figure*}
\centering
\includegraphics[width=0.33\columnwidth]{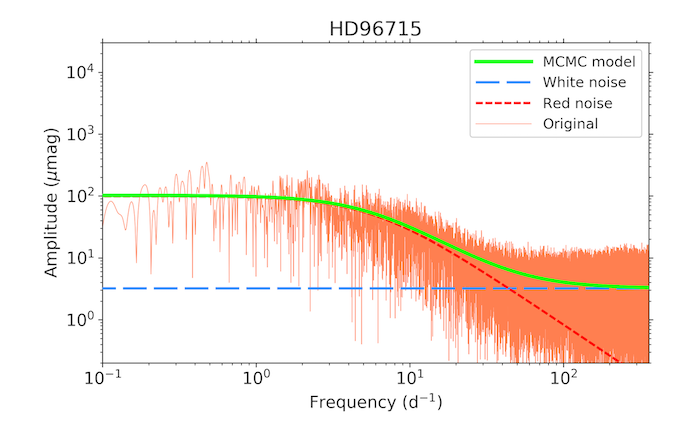}				
\includegraphics[width=0.33\columnwidth]{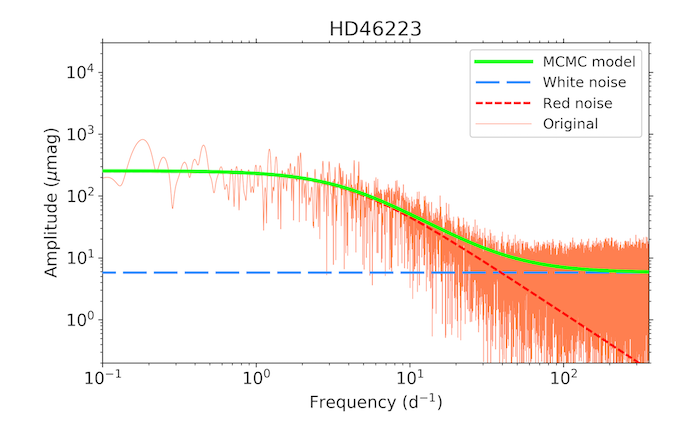}				
\includegraphics[width=0.33\columnwidth]{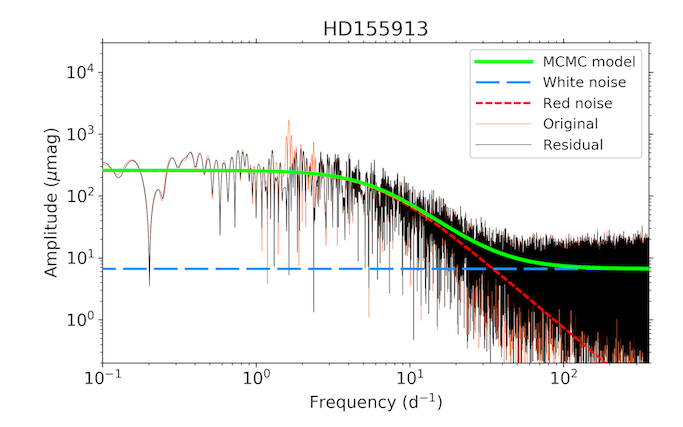}			
\includegraphics[width=0.33\columnwidth]{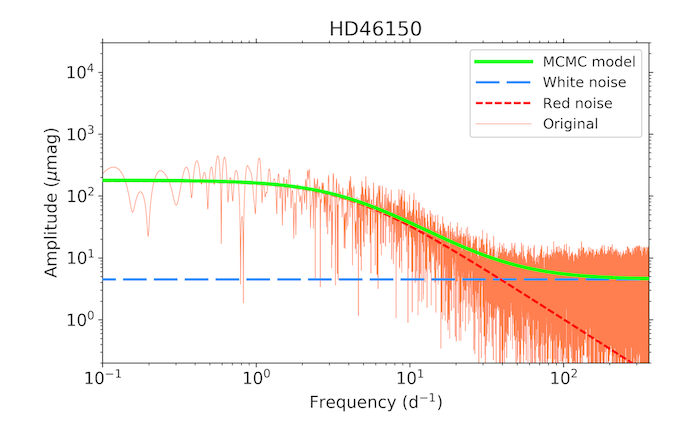}				
\includegraphics[width=0.33\columnwidth]{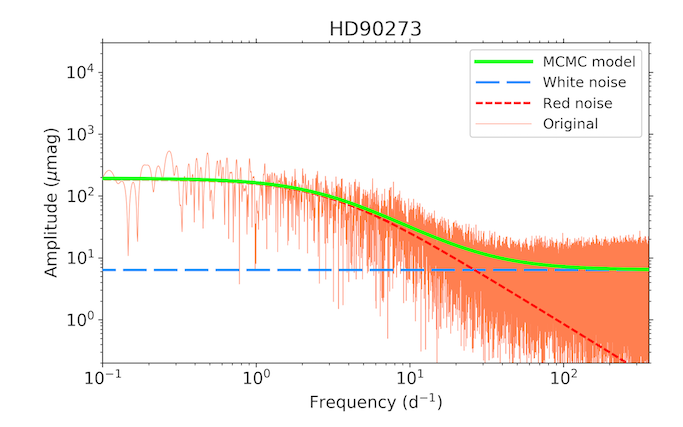}				
\includegraphics[width=0.33\columnwidth]{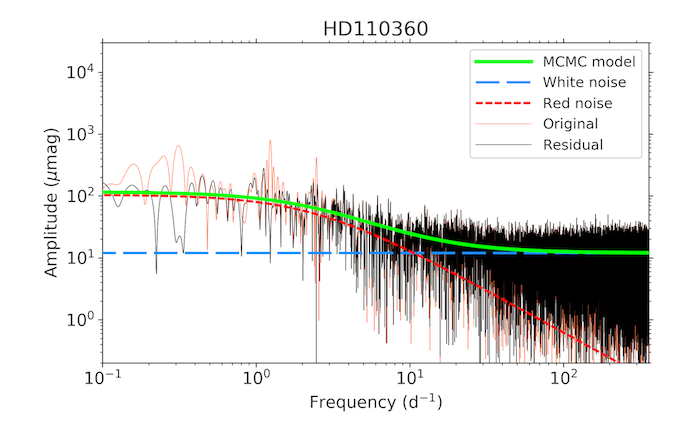}			
\includegraphics[width=0.33\columnwidth]{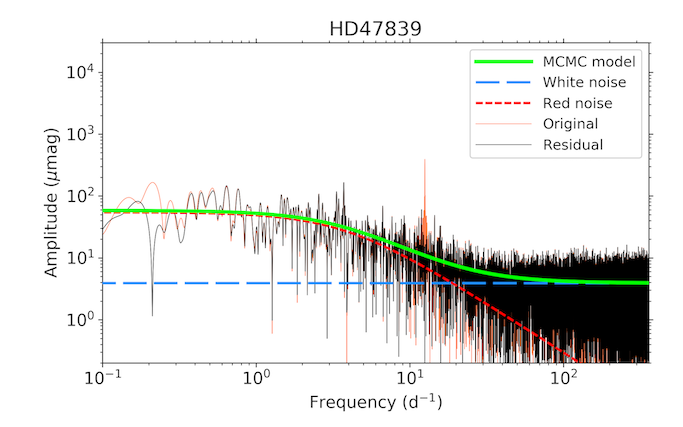}			
\includegraphics[width=0.33\columnwidth]{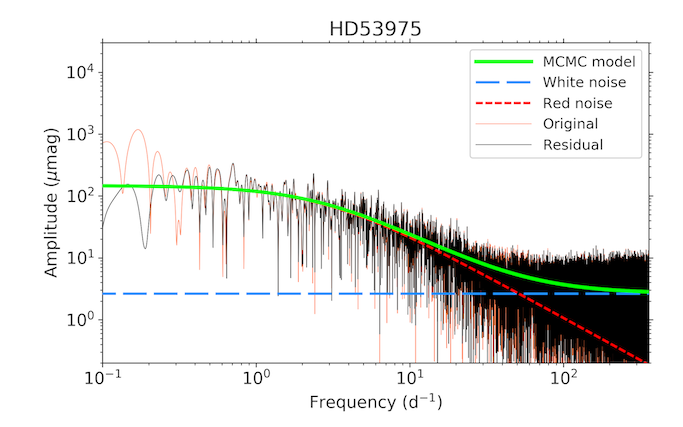}			
\includegraphics[width=0.33\columnwidth]{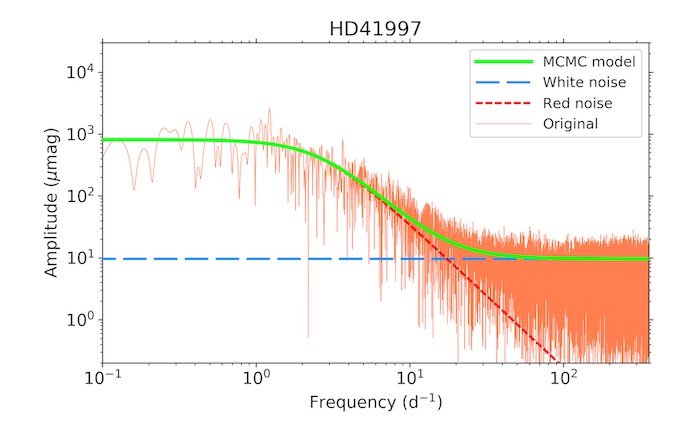}				
\includegraphics[width=0.33\columnwidth]{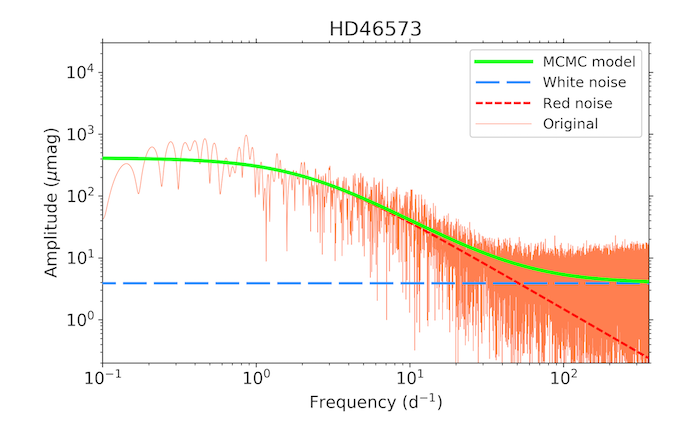}				
\includegraphics[width=0.33\columnwidth]{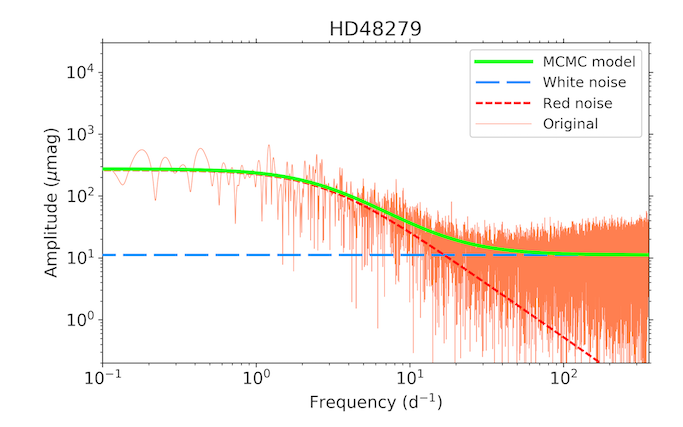}				
\includegraphics[width=0.33\columnwidth]{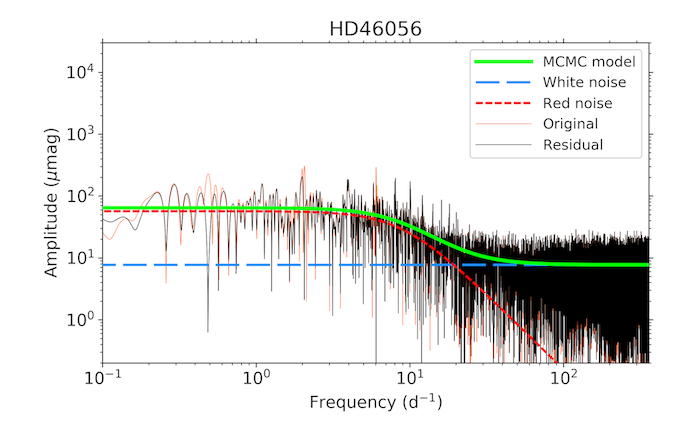}			
\includegraphics[width=0.33\columnwidth]{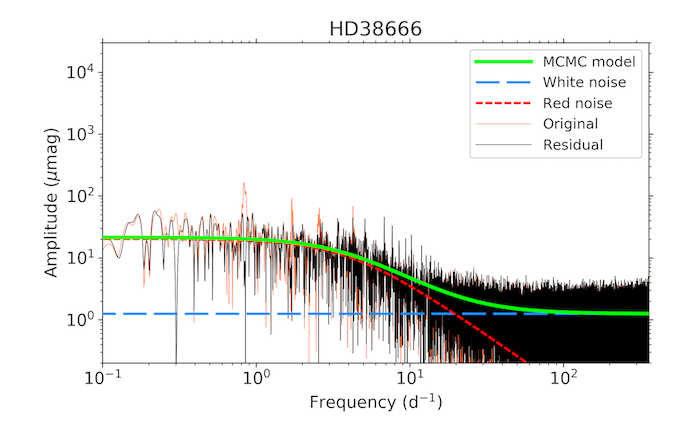}			
\includegraphics[width=0.33\columnwidth]{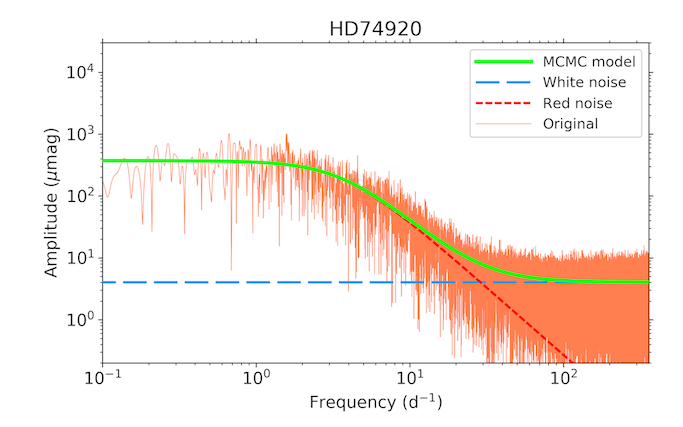}				
\includegraphics[width=0.33\columnwidth]{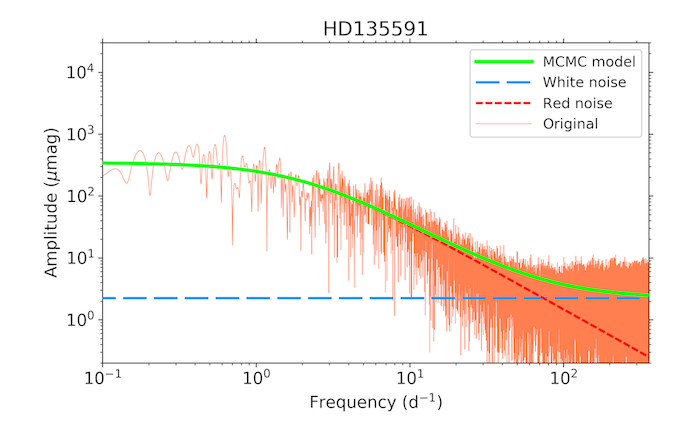}				
\includegraphics[width=0.33\columnwidth]{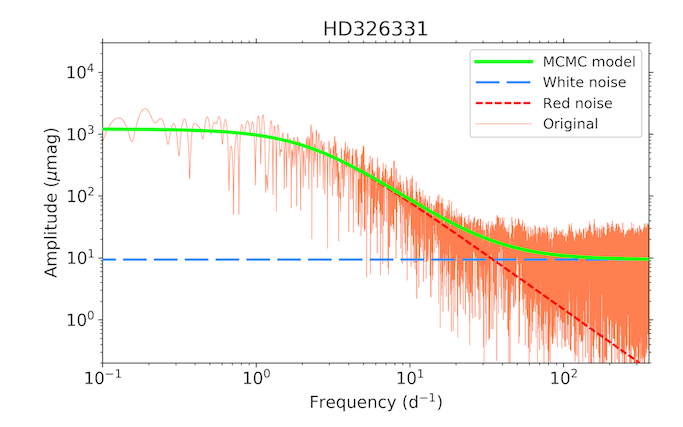}				
\includegraphics[width=0.33\columnwidth]{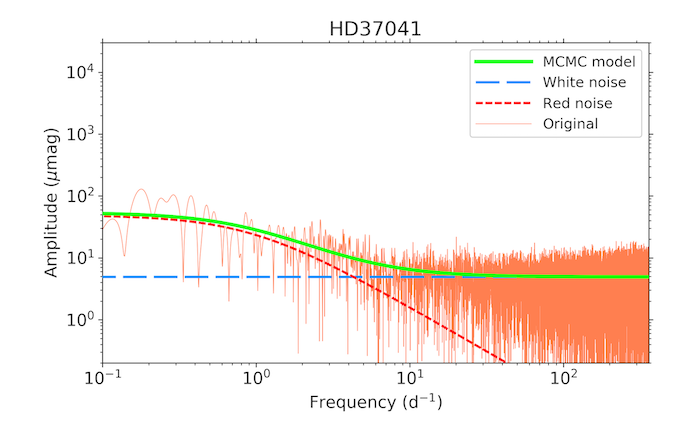}				
\includegraphics[width=0.33\columnwidth]{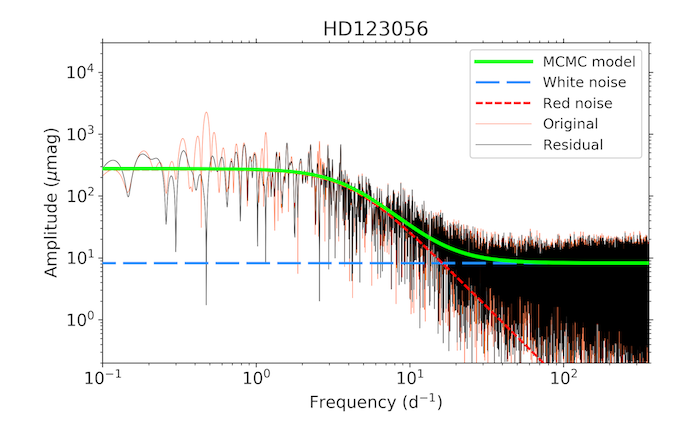}			
\caption{Fitted logarithmic amplitude spectra of stars given in Table~\ref{table: params}. Line styles and colours are the same as in Fig.~\ref{figure: TESS}.}
\label{figure: B1}
\end{figure*}



\begin{figure*}
\centering
\includegraphics[width=0.33\columnwidth]{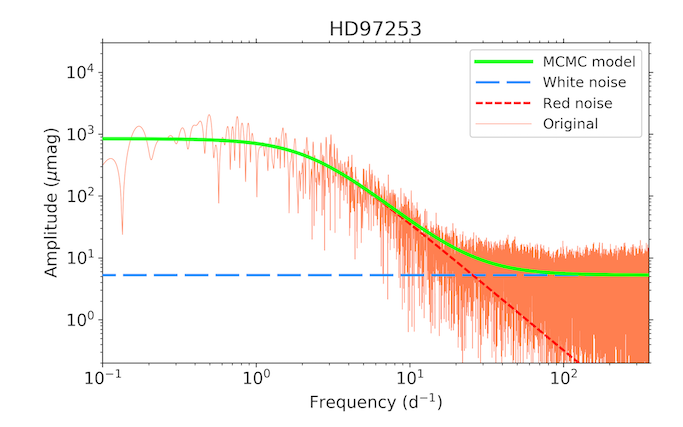}				
\includegraphics[width=0.33\columnwidth]{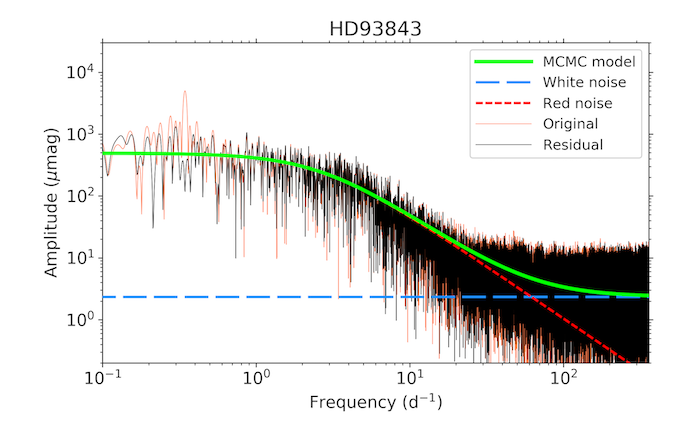}			
\includegraphics[width=0.33\columnwidth]{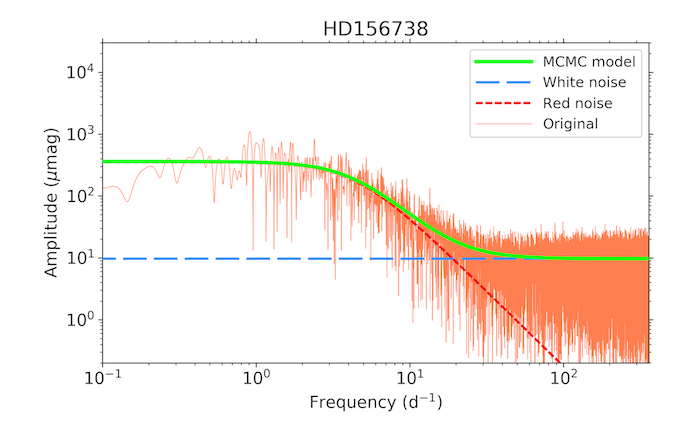}				
\includegraphics[width=0.33\columnwidth]{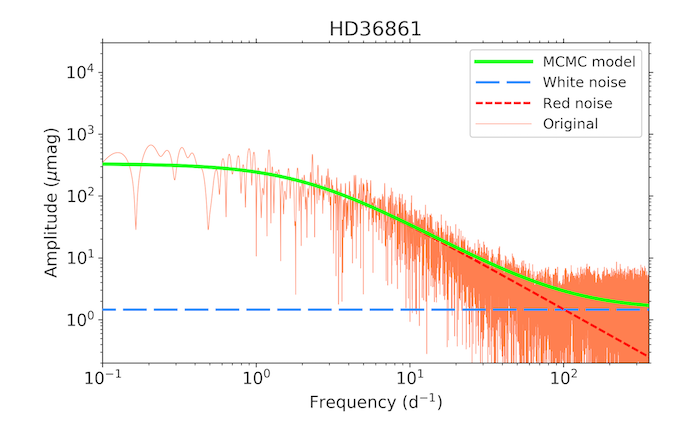}				
\includegraphics[width=0.33\columnwidth]{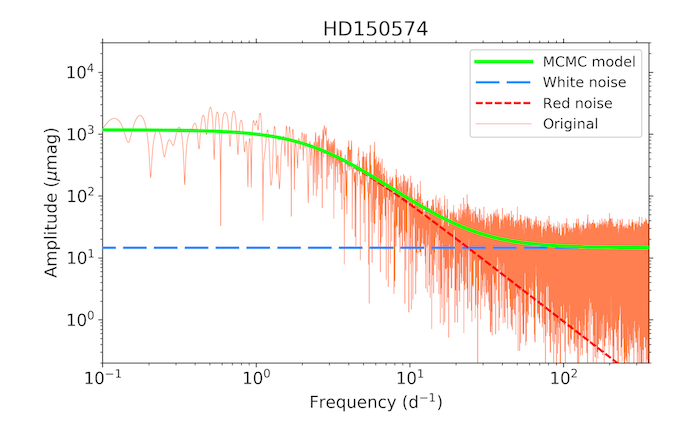}				
\includegraphics[width=0.33\columnwidth]{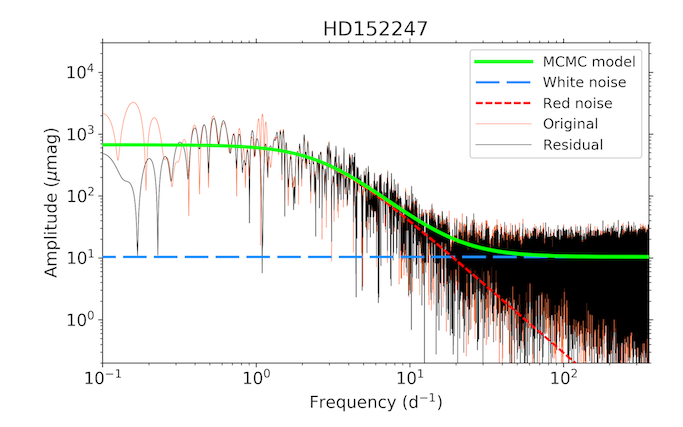}			
\includegraphics[width=0.33\columnwidth]{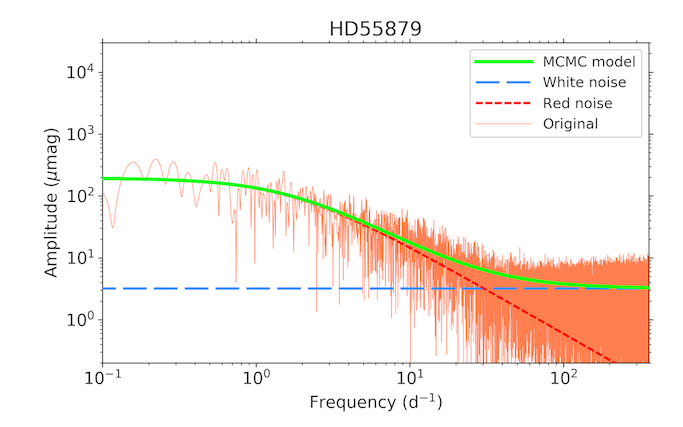}				
\includegraphics[width=0.33\columnwidth]{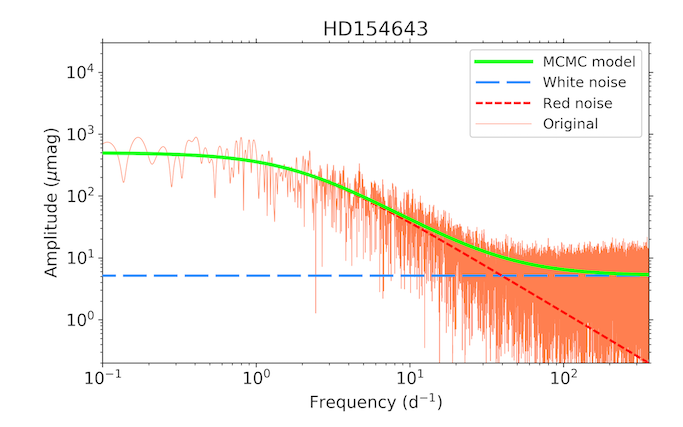}				
\includegraphics[width=0.33\columnwidth]{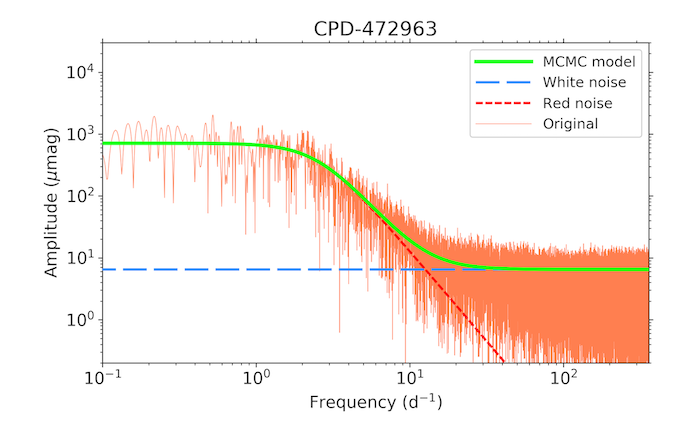}				
\includegraphics[width=0.33\columnwidth]{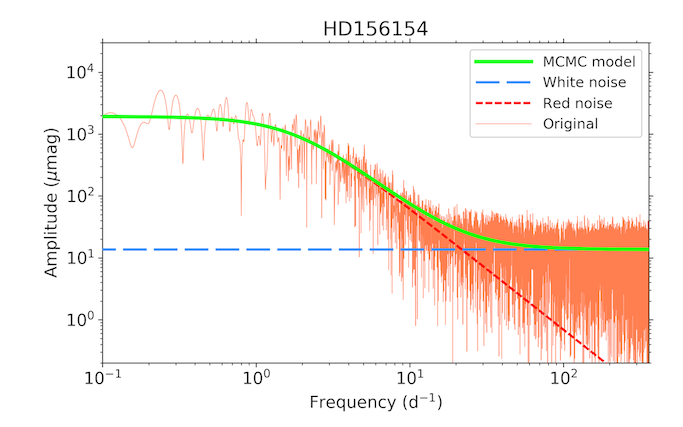}				
\includegraphics[width=0.33\columnwidth]{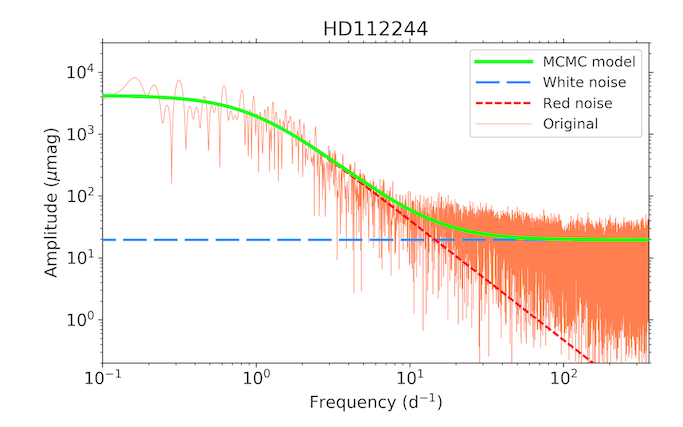}				
\includegraphics[width=0.33\columnwidth]{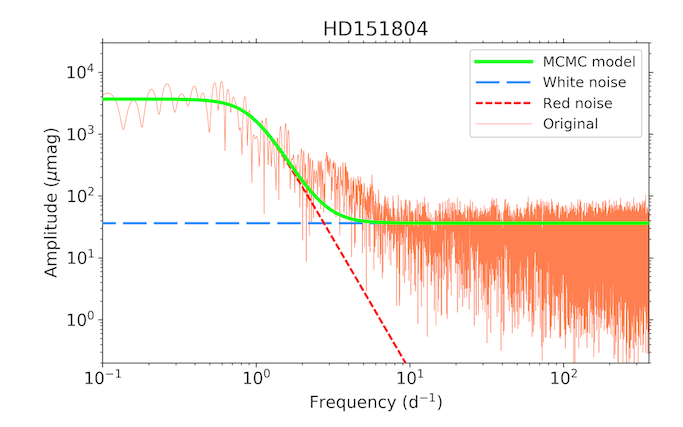}				
\includegraphics[width=0.33\columnwidth]{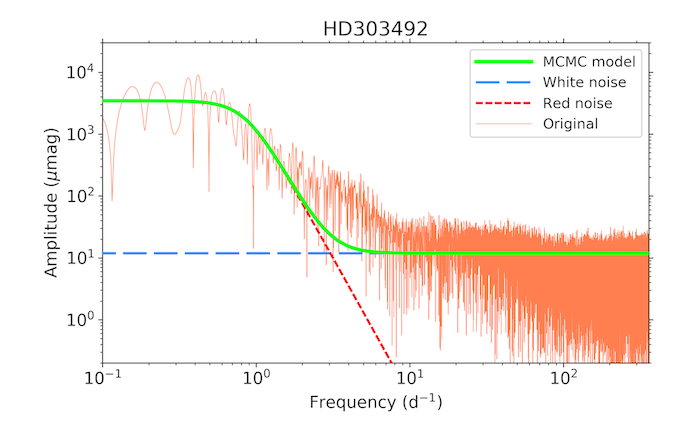}				
\includegraphics[width=0.33\columnwidth]{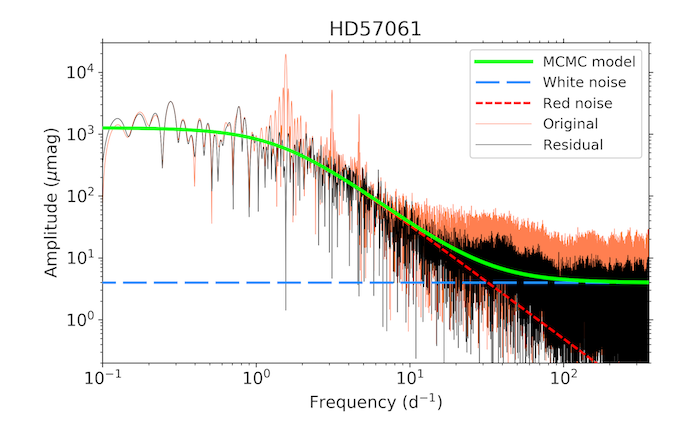}		
\includegraphics[width=0.33\columnwidth]{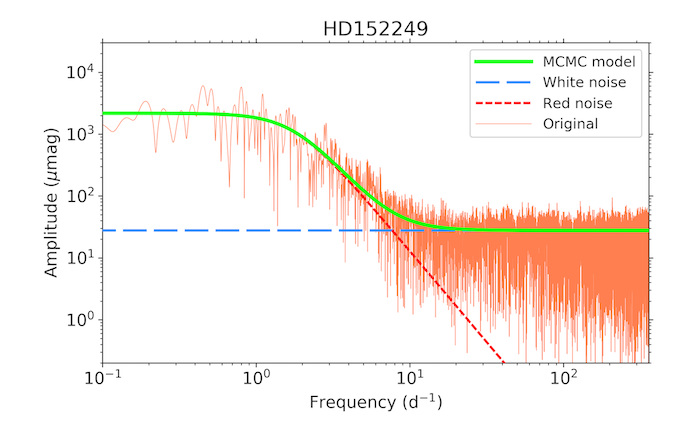}				
\includegraphics[width=0.33\columnwidth]{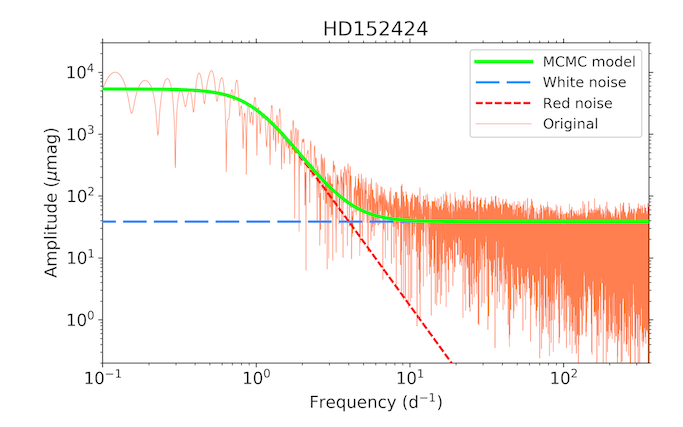}				
\includegraphics[width=0.33\columnwidth]{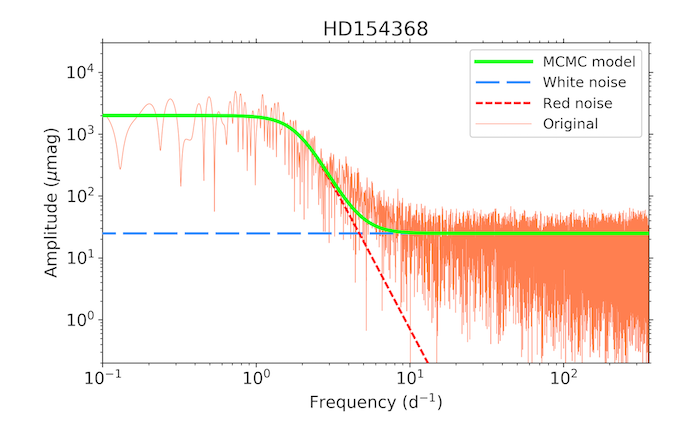}				
\includegraphics[width=0.33\columnwidth]{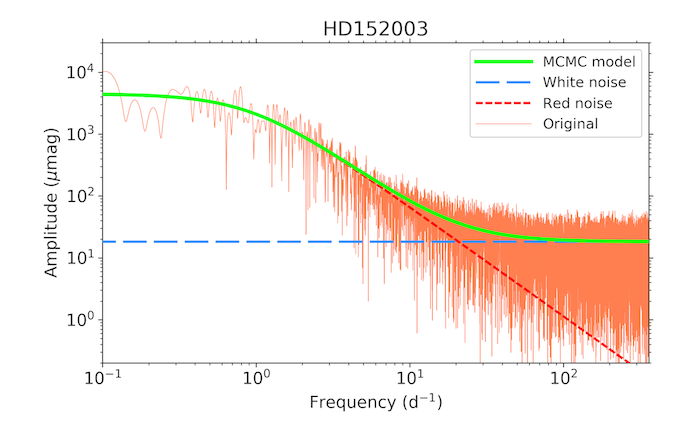}				
\caption{Fitted logarithmic amplitude spectra of stars given in Table~\ref{table: params}. Line styles and colours are the same as in Fig.~\ref{figure: TESS}.}
\label{figure: B2}
\end{figure*}


\begin{figure*}
\centering
\includegraphics[width=0.33\columnwidth]{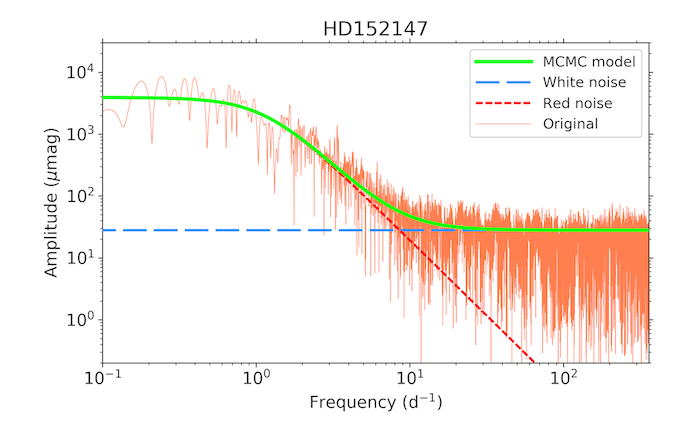}				
\includegraphics[width=0.33\columnwidth]{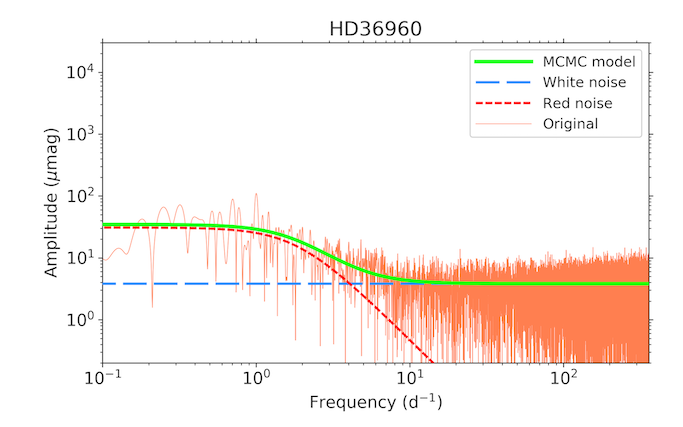}				
\includegraphics[width=0.33\columnwidth]{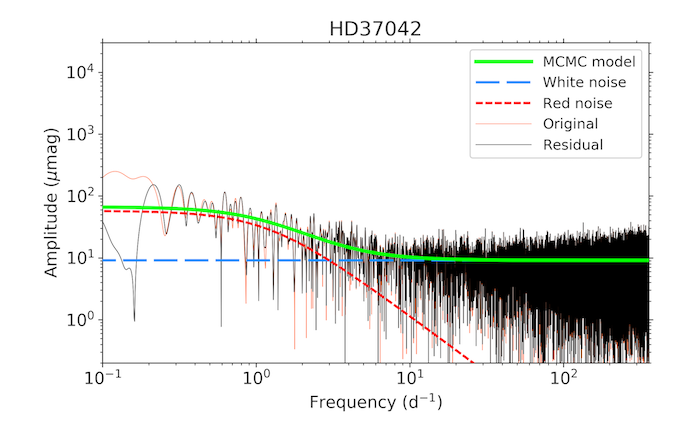}			
\includegraphics[width=0.33\columnwidth]{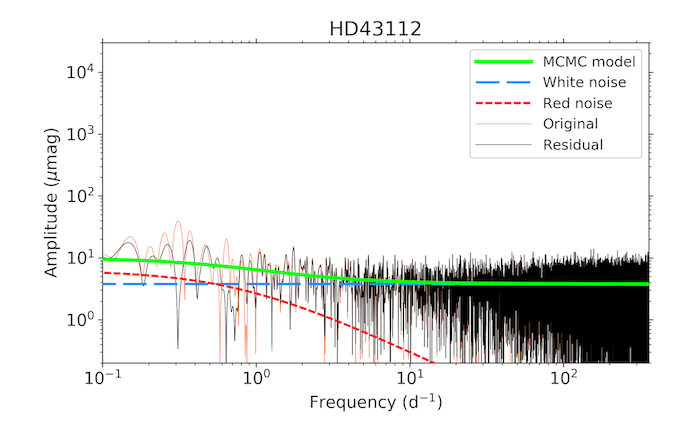}			
\includegraphics[width=0.33\columnwidth]{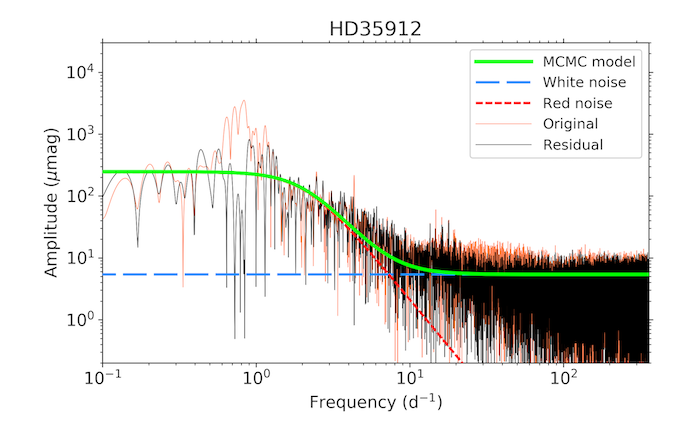}		
\includegraphics[width=0.33\columnwidth]{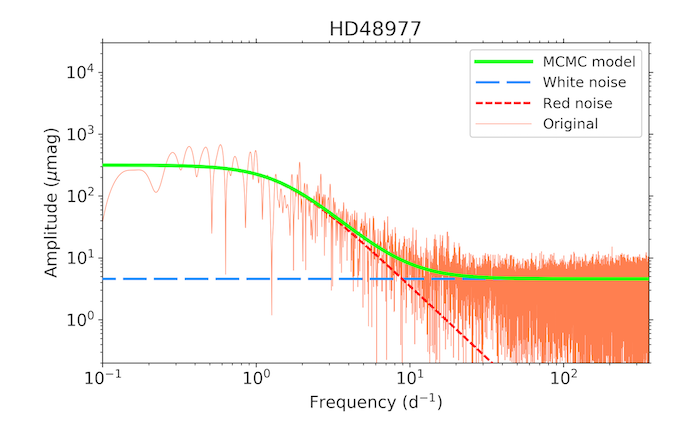}		
\includegraphics[width=0.33\columnwidth]{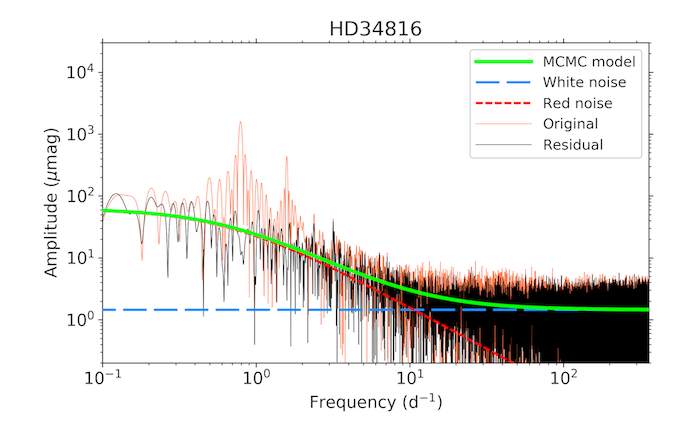}			
\includegraphics[width=0.33\columnwidth]{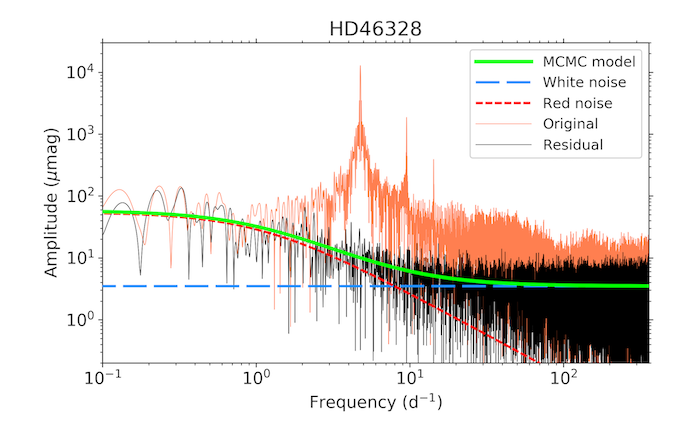}		
\includegraphics[width=0.33\columnwidth]{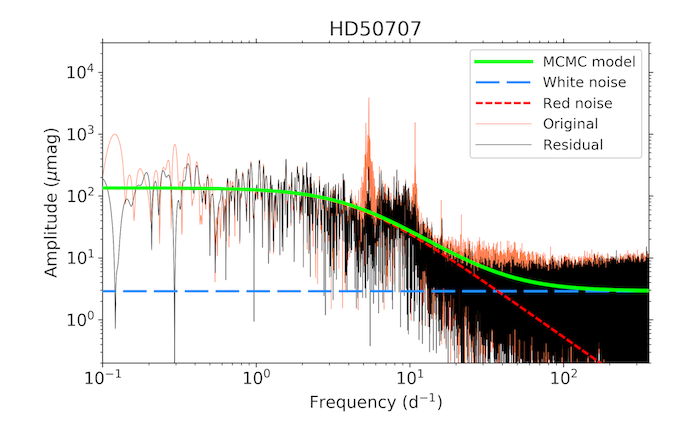}		
\includegraphics[width=0.33\columnwidth]{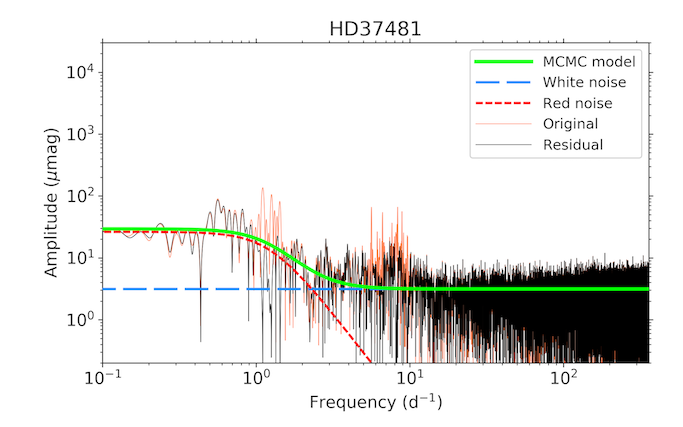}		
\includegraphics[width=0.33\columnwidth]{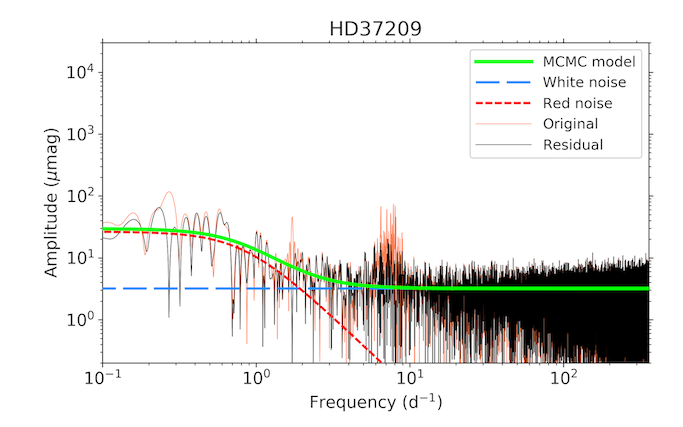}		
\includegraphics[width=0.33\columnwidth]{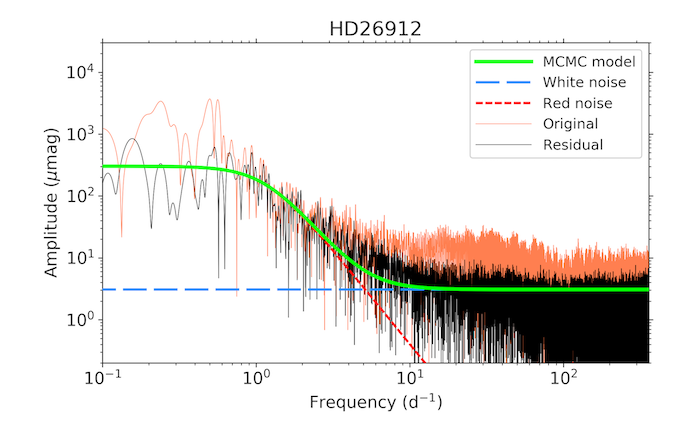}			
\includegraphics[width=0.33\columnwidth]{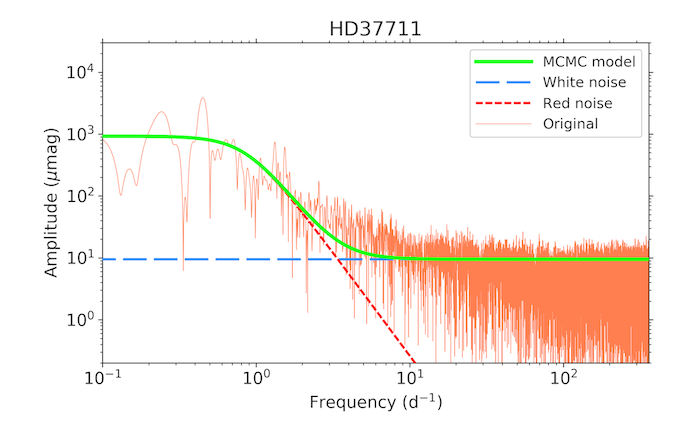}				
\includegraphics[width=0.33\columnwidth]{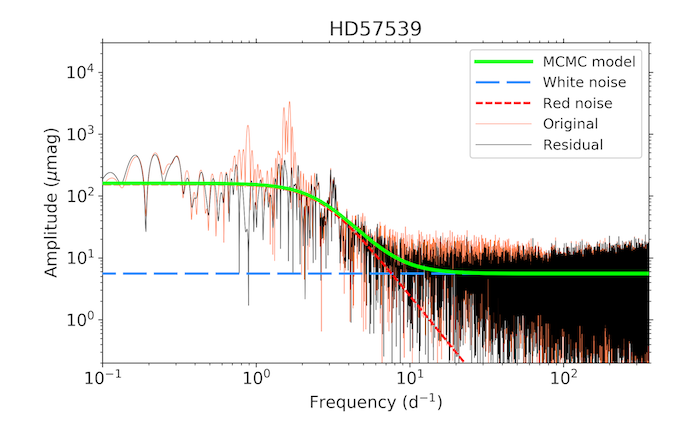}			
\includegraphics[width=0.33\columnwidth]{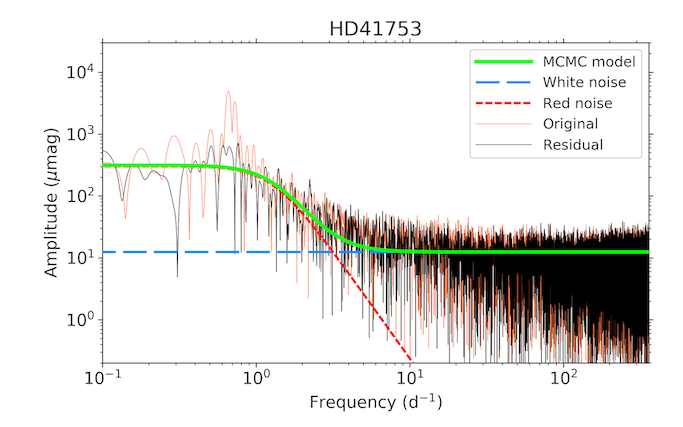}			
\includegraphics[width=0.33\columnwidth]{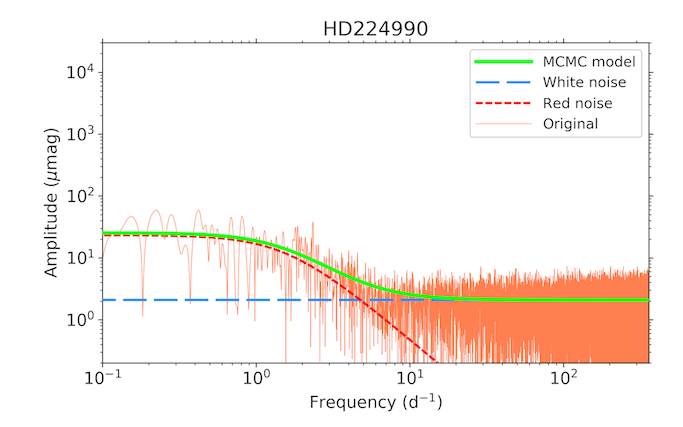}			
\includegraphics[width=0.33\columnwidth]{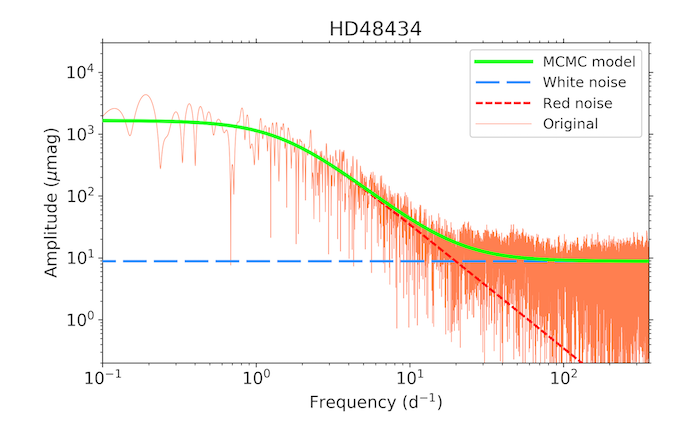}				
\includegraphics[width=0.33\columnwidth]{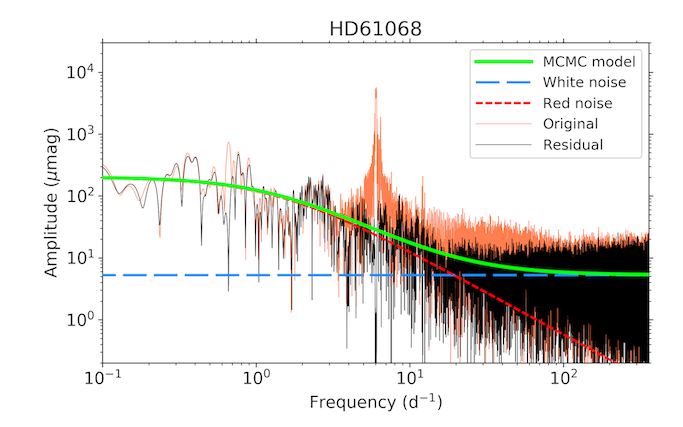}		
\caption{Fitted logarithmic amplitude spectra of stars given in Table~\ref{table: params}. Line styles and colours are the same as in Fig.~\ref{figure: TESS}.}
\label{figure: B3}
\end{figure*}


\begin{figure*}
\centering
\includegraphics[width=0.33\columnwidth]{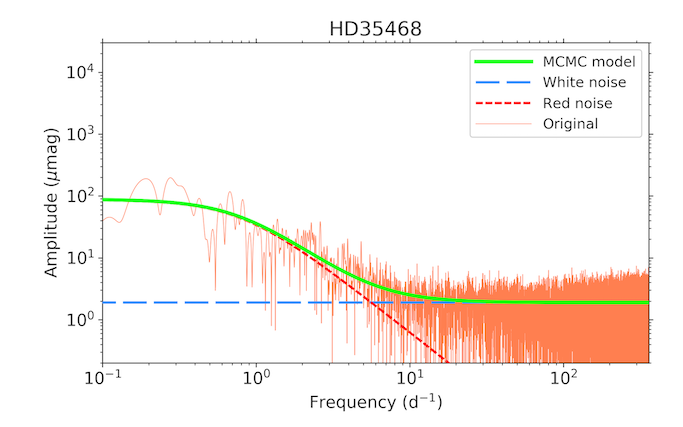}				
\includegraphics[width=0.33\columnwidth]{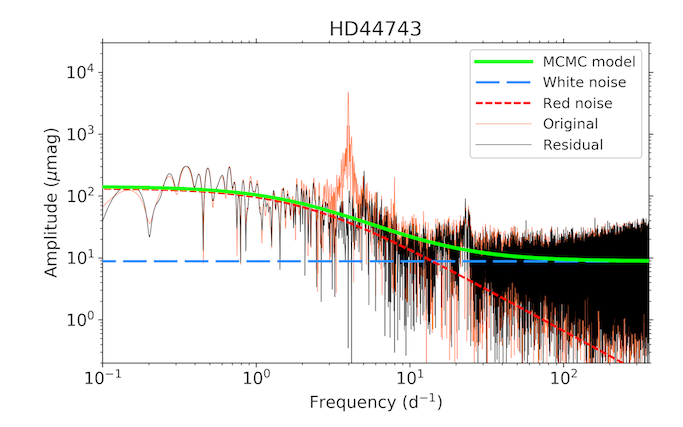}		
\includegraphics[width=0.33\columnwidth]{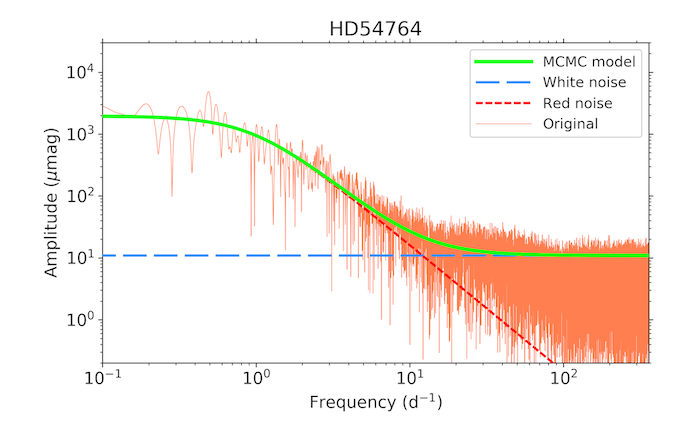}				
\includegraphics[width=0.33\columnwidth]{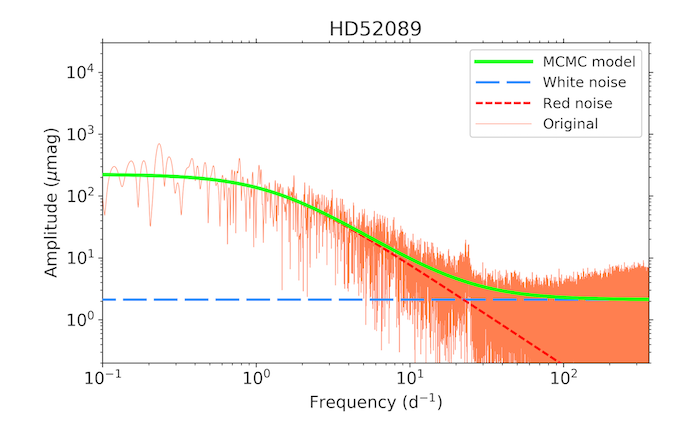}				
\includegraphics[width=0.33\columnwidth]{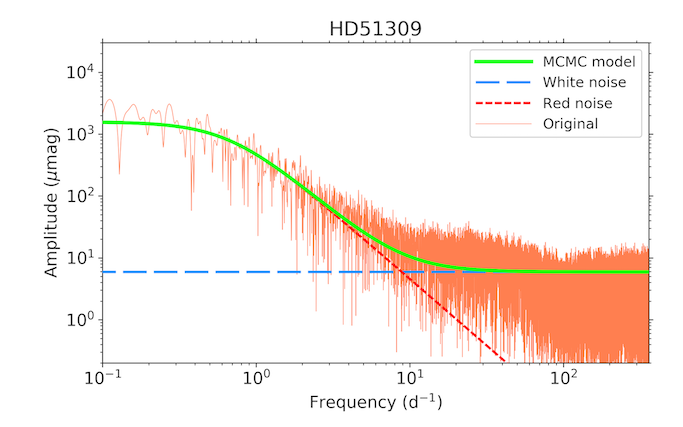}				
\includegraphics[width=0.33\columnwidth]{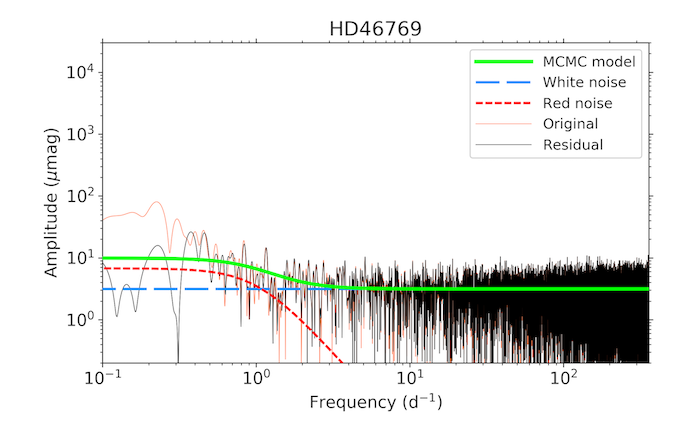}			
\includegraphics[width=0.33\columnwidth]{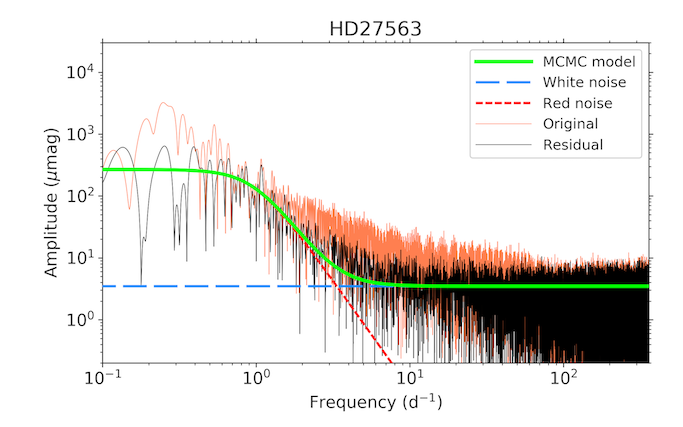}			
\includegraphics[width=0.33\columnwidth]{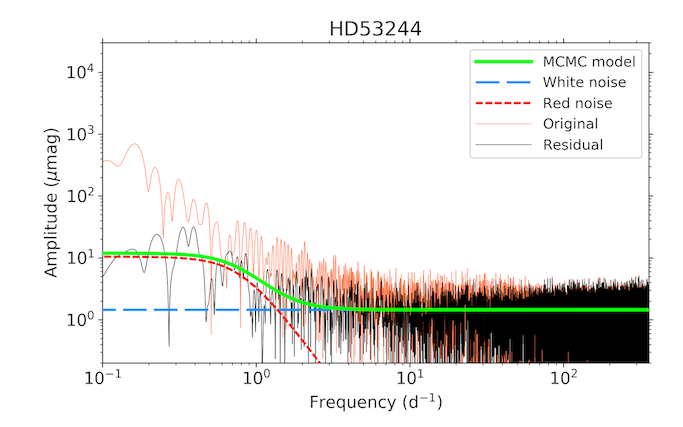}			
\includegraphics[width=0.33\columnwidth]{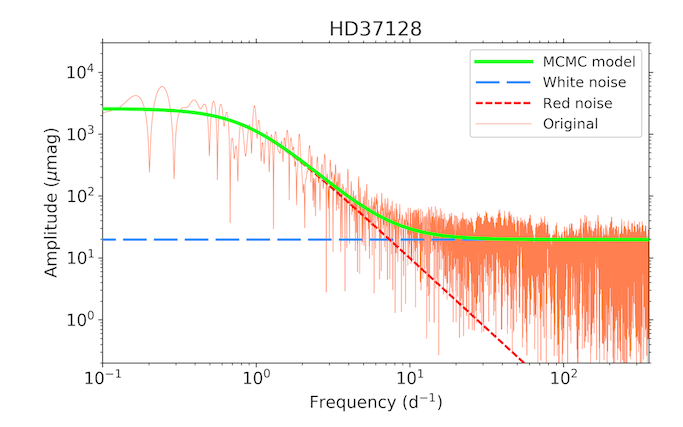}				
\includegraphics[width=0.33\columnwidth]{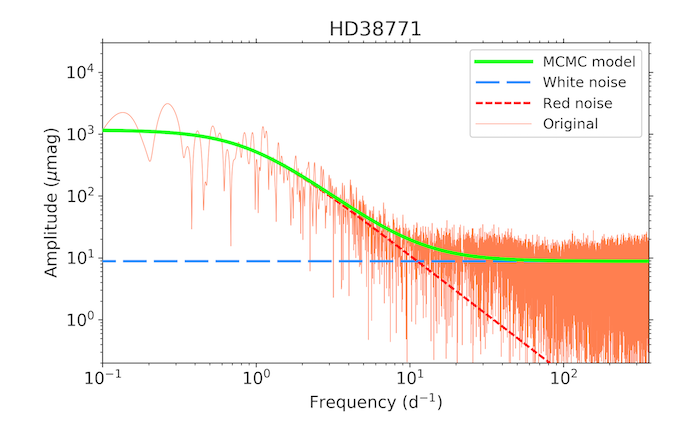}				
\includegraphics[width=0.33\columnwidth]{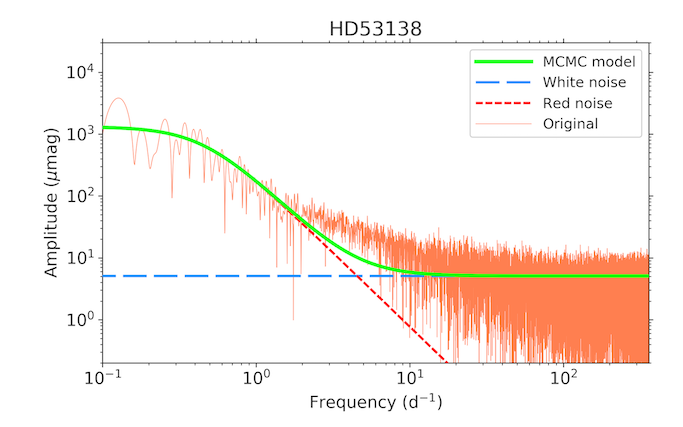}				
\includegraphics[width=0.33\columnwidth]{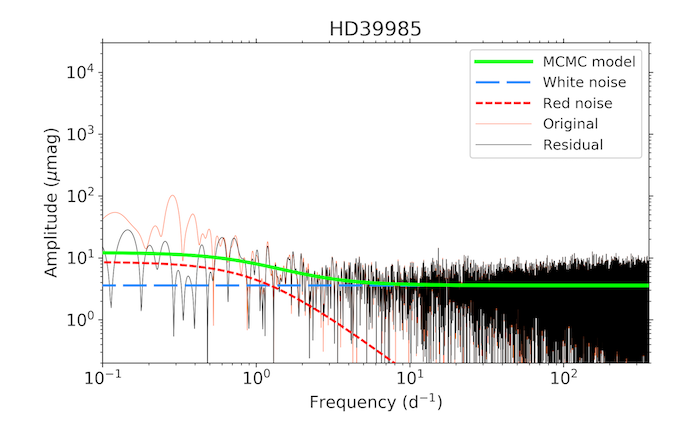}			
\includegraphics[width=0.33\columnwidth]{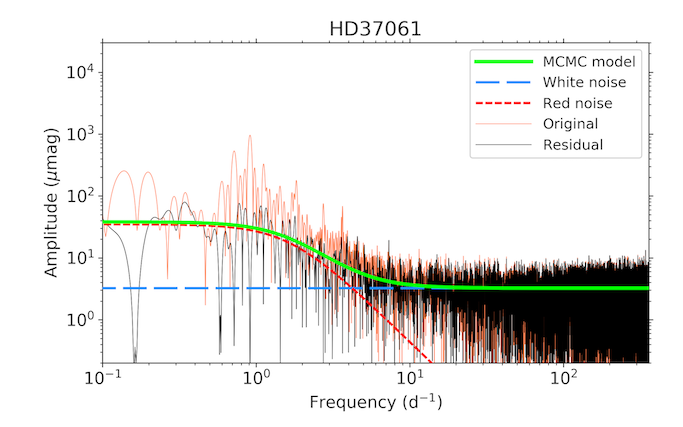}		
\includegraphics[width=0.33\columnwidth]{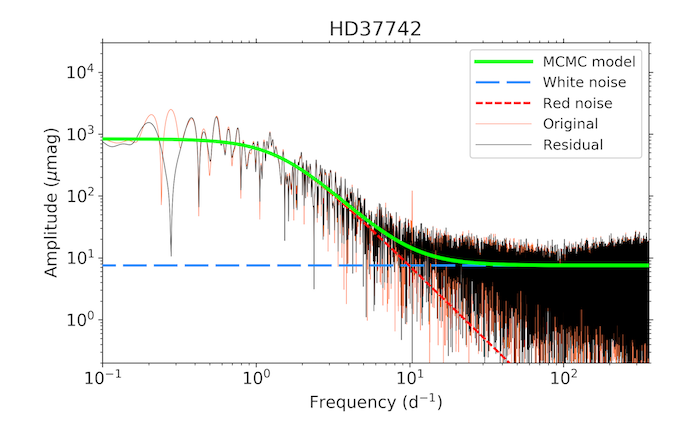}		
\includegraphics[width=0.33\columnwidth]{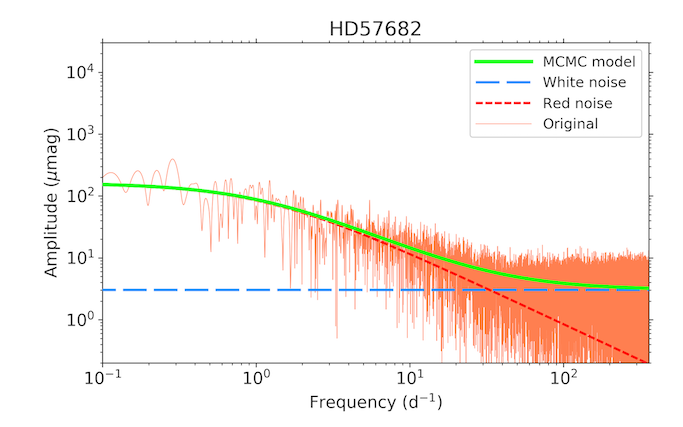}				
\includegraphics[width=0.33\columnwidth]{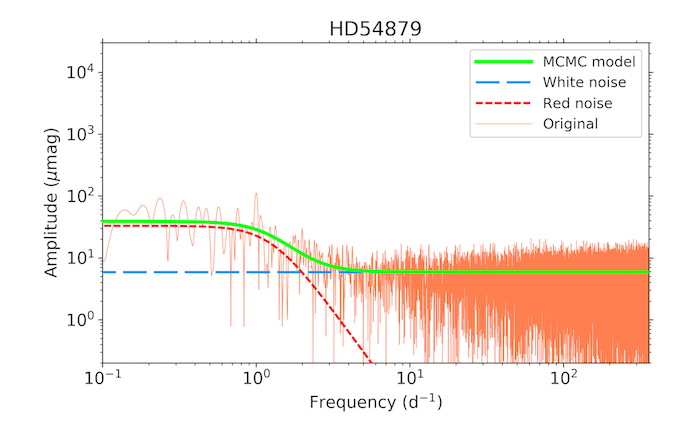}				
\caption{Fitted logarithmic amplitude spectra of stars given in Table~\ref{table: params}. Line styles and colours are the same as in Fig.~\ref{figure: TESS}.}
\label{figure: B4}
\end{figure*}



\end{appendix}


\end{document}